\pdfoutput=1
\documentclass[aps,prd,amsmath,floats,floatfix, twocolumn,
superscriptaddress,nofootinbib,showpacs]{revtex4-2}

\usepackage[T1]{fontenc}
\usepackage[utf8]{inputenc}
\usepackage{lmodern}
\usepackage{verbatim}
\usepackage[dvipsnames, usenames]{xcolor}
\definecolor{linkcolor}{rgb}{0.0,0.3,0.5}
\usepackage[hypertexnames=false, unicode, colorlinks=true, linkcolor=linkcolor,
citecolor=linkcolor, filecolor=linkcolor,urlcolor=linkcolor, pdfusetitle]{hyperref}

\usepackage[all]{hypcap}
\usepackage{graphicx}
\usepackage{xspace}
\usepackage{amssymb}
\usepackage[normalem]{ulem} 
\usepackage{bm} 
\usepackage{enumitem,amssymb}
\usepackage{orcidlink}
\usepackage{caption}
\usepackage{multirow}

\usepackage{microtype}
\usepackage[english]{babel}
\usepackage{blindtext}

\graphicspath{%
  {figs/}%
}

\DeclareMathAlphabet{\mathpzc}{OT1}{pzc}{m}{it}

\newlist{todolist}{itemize}{2}
\setlist[todolist]{label=$\square$}

\newcommand{\bea}{\begin{eqnarray}}
\newcommand{\eea}{\end{eqnarray}}
\newcommand{\be}{\begin{equation}}
\newcommand{\ee}{\end{equation}}

\newcommand{\boldtheta}{\boldsymbol{\theta}}

\definecolor{dkgreen}{rgb}{0,0.6,0}
\definecolor{gray}{rgb}{0.5,0.5,0.5}
\definecolor{mauve}{rgb}{0.58,0,0.82}

\newcommand{\bet}{\ensuremath{\frac{96}{5}\pi^{8/3}\left(\frac{G}{c^3}\right)^{5/3}M^{5/3}_cf^{8/3}_0}}

\newcommand\anu{\affiliation{OzGrav-ANU, Centre for Gravitational Astrophysics, Research School of Physics and Research School of Astronomy \& Astrophysics, The Australian National University, ACT 2601, Australia}}
\newcommand\INFNRoma{\affiliation{INFN, Sezione di Roma, I-00185 Roma, Italy}}
\newcommand\uib{\affiliation{IAC3–IEEC, Universitat de les Illes Balears, E-07122 Palma de Mallorca, Spain}}

\begin{document}

\title{A semicoherent resampling method for long-transient gravitational wave searches with applications to subsolar-mass primordial black-hole binaries}

\author{Neil Lu \orcidlink{0000-0002-8861-9902}}
\email{neil.lu@anu.edu.au }
\anu

\author{Cristiano Palomba}
\INFNRoma

\author{Ornella J. Piccinni \orcidlink{0000-0001-5478-3950}}
\uib

\author{Ling Sun \orcidlink{0000-0001-7959-892X}}
\anu

\date{\today}

\begin{abstract}
Resampling removes the modeled phase evolution of a frequency-evolving signal, transforming it into a monochromatic one. 
We introduce a novel implementation for long-transient gravitational-wave searches that evaluates the Fourier spectrum of the resampled data using a type-I non-uniform fast Fourier transform (NUFFT). This implementation can be up to two orders of magnitude faster than some previous implementations, depending on coherent length, sampling rate, and desired accuracy.
To demonstrate its application, we benchmark the method with a semicoherent search for long-duration inspiral signals from subsolar-mass
primordial black hole binaries. For a search over $40$--$60\,\mathrm{Hz}$ and chirp masses of $5\times10^{-4}$--$10^{-1}\,M_\odot$, with a coherent duration of $30\,\mathrm{s}$ and a realistic computing cost, the estimated horizon distance exceeds the Galactic Center for all of the parameter space considered and exceeds Andromeda for chirp masses above approximately $2\times10^{-2}M_\odot$. 
These results establish NUFFT-based resampling as a computationally efficient and broadly applicable approach for searches for modeled, frequency-evolving gravitational-wave signals.
\end{abstract}

\maketitle


\section{Introduction}
\label{sec:introduction}

A decade after the first direct detection of gravitational waves \cite{LIGOScientific:2016aoc}, gravitational-wave astronomy has become an established observational probe of compact-object astrophysics and strong-field gravity. The LIGO \cite{LIGOScientific:2014pky}, Virgo \cite{VIRGO:2014yos}, and KAGRA \cite{KAGRA:2020tym} detectors have detected hundreds of compact-binary coalescence signals, transforming the original discovery into a growing population of astrophysical sources \cite{LIGOScientific:2026wfs}. The currently observed catalog consists of compact binary mergers whose signal durations in the sensitivity band of ground-based interferometers are typically short, ranging from fractions to tens of seconds.

There is growing interest in transient gravitational-wave signals that remain detectable for much longer times. These could be produced by sources including post-merger remnants of neutron-star mergers \cite{LIGOScientific:2017fdd, Grace:2023kqq}, pulsar glitches or their aftermath \cite{Keitel:2019zhb, Haskell:2023exob}, low-mass primordial black hole (PBH) binaries \cite{Miller:2024rca, Andres-Carcasona:2024jvz, Rodriguez:2026oyia}, and ultralight vector boson clouds around black holes \cite{Siemonsen:2019ebd, Jones:2023fzz, LIGOScientific:2025csr}. These signals offer substantial scientific opportunities, but their extended duration compared to transient signals also creates difficulties in the analysis. The template-bank size and computational cost of matched filtering, the conventionally used technique for detecting signals, can increase rapidly with the signal duration \cite{Owen:1998dk}.

A natural way to reduce this cost is to use semicoherent methods. Such methods are standard in searches for continuous gravitational waves (also referred to as continuous waves), where fully coherent integration over months or years is often computationally prohibitive. In a semicoherent search, the data are divided into shorter segments, a detection statistic is coherently computed in each segment, and the information from different segments is then combined incoherently; see e.g., Refs.~\cite{Wette:2023dom, Tenorio:2021wmz} for reviews of these methods. The incoherent summation between segments sacrifices phase coherence and reduces sensitivity, but can dramatically reduce computational cost and enable broader searches over otherwise inaccessible parameter spaces. Semicoherent methods have been used to search for both modeled \cite{Piccinni:2018akm, Prix:2011qv,Suvorova:2016rdc, Sun:2017zge} and unmodeled \cite{Sun:2018owi, Macquet:2021ttqa,Banagiri:2019obu} signals. 

Semicoherent methods generally require the data to remain confined within a single Fourier bin in each segment. For slowly evolving continuous-wave signals, such as those from rotating neutron stars, this condition can be satisfied even for long coherence times as the frequency evolution is often slow. For more rapidly evolving signals, however, the frequency evolution would demand short coherence lengths to ensure the signal remains confined to a single Fourier bin (whose width is the inverse of the coherence time), causing loss of sensitivity. This motivates a preprocessing step known as resampling that removes the expected frequency evolution before constructing the spectra for each segment. Given a known signal model, the time series can be resampled to remove the modeled frequency evolution, after which it becomes monochromatic and remains confined within a Fourier bin. This enables semicoherent analysis with a longer coherence duration and greater sensitivity. 

For searches over signal durations from $\mathcal{O}(10^2\,\mathrm{s})$ to $\mathcal{O}(10\,\mathrm{months})$, the computational cost is, as with continuous-wave analysis, dominated by the generation of the resampled time series and the subsequent Fourier transforms. In this work, we present a resampling implementation which accelerates both the frequency correction step and the subsequent production of Fourier spectra. We demonstrate the applicability of this technique in searches for binary subsolar-mass black holes (BHs), potentially of primordial origin, which may have formed in the early Universe \cite{Hawking:1971ei, Green:2020jor, Carr:2023tpt, LISACosmologyWorkingGroup:2023njw}. Unlike astrophysical BHs formed through stellar collapse, PBHs could exist with subsolar masses \cite{Carr:1975qj} and therefore emit signals with durations lasting from tens of seconds to years. Within the sensitive range of audio-band gravitational-wave detectors, the signals from these systems would remain in their inspiral phase, with the merger occurring at higher frequencies. 

The structure of the paper is as follows. In Sec.~\ref{sec:signal_model}, we describe the signal model for subsolar-mass BH inspirals and review their basic properties. In Sec.~\ref{sec:methods}, we detail a semicoherent approach which can be used to search for the subsolar-mass BH inspirals and provide details on the use of NUFFTs in the search. In Sec.~\ref{sec:distance_sens}, we analytically derive the horizon distance this semicoherent method could reach. In Sec.~\ref{sec:injection}, we verify the analysis procedure with synthetic signals injected into detector noise. We discuss the computational cost of the semicoherent search in Sec.~\ref{sec:cost}. Finally, we conclude and discuss future work in Sec.~\ref{sec:conclusion}. 

\section{Signal model}
\label{sec:signal_model}
In this work, we consider the inspiral of subsolar-mass PBH binaries as a specific type of source for the resampling approach. Binaries with total masses less than $0.1 M_\odot$ remain in the inspiral regime within the sensitivity band of ground-based detectors and coalesce at higher frequencies. The boundary of this regime can be taken as the gravitational-wave emission frequency at the innermost stable circular orbit (ISCO), i.e., the smallest radius at which a stable circular orbit is possible, beyond which the signal has definitively exited the inspiral regime:
\begin{gather}
    f_{\rm ISCO} \simeq 4400 \frac{M_\odot}{M_{\rm tot}} \, \mathrm{Hz} \, ,
\end{gather}
where $M_{\rm tot}$ denotes the total mass of the binary system. 

It is also useful to consider the signal durations of subsolar-mass PBHs. The inspiral evolution of a binary BH system is described by Post-Newtonian (PN) theory, a perturbative expansion in powers of the characteristic orbital velocity $v/c$, where $c$ is the speed of light. Using the leading-order term (0PN) allows us to estimate the time taken for a system to evolve from a starting frequency $f_0$ to a final frequency $f_{\rm end}$, given by:
\begin{gather}
    t_{\rm 0 \rightarrow end} \simeq \frac{3}{8\beta(f_0, M_c)}\left( 1-\left(\frac{f_0}{f_{\rm end}}\right)^{8/3}\right) \, , \\
    \beta(f_0, M_c) = \bet \, , \label{eq:beta}
\end{gather}
where $G$ is the gravitational constant, \mbox{$M_c = {(m_1 m_2)^{3/5}}{(m_1 + m_2)^{-1/5}}$} is the chirp mass of the binary system, $m_1$ and $m_2$ are the masses of the two constituent BHs, and $f_0$ is the frequency at a reference time $t_0$. 
For a system with a chirp mass of $10^{-1}M_\odot$ ($5 \times 10^{-4}M_\odot$), it takes approximately 3 hours (2 years) to chirp across the audio-detector band of 20--2000 Hz. The signal morphology of subsolar-mass PBHs is therefore distinct from both standard compact-binary-coalescence (CBC) and continuous-wave signals. Compared to stellar-mass CBCs, the signal evolution evolved more slowly, sorequires a longer coherent integration time. However, unlike continuous waves, the signal is not well approximated as monochromatic over the full observation time. The search problem is therefore best viewed as a long-duration transient search, rather than as either a short transient CBC search or a nearly monochromatic continuous-wave search.

We construct waveforms at the 3.5PN order \cite{Blanchet:2001ax, Faye:2012we} and restrict our analysis to quasi-circular, non-spinning, equal-mass systems, such that the waveform is parameterized by $\{f_0, M_c\}$. This PN order was, until recently, the highest for which a complete analytical expression for the inspiral phasing was available, although the quasi-circular, non-spinning phase has now been extended to the 4.5PN order \cite{Blanchet:2023bwj}. We expect the higher-order terms beyond the 3.5PN order to have a negligible effect in the parameter space considered here: for an equal-mass binary with $M_c=1.2M_\odot$, the combined 4PN and 4.5PN contributions amount to only approximately $0.01$ gravitational-wave cycles when integrated from 30--1000 Hz \cite{Blanchet:2023bwj}. The lower-mass systems targeted by this search remain much farther from their ISCOs over the analysis band, where the higher-order PN terms accumulate substantially less dephasing. We use TaylorF2 \cite{Buonanno:2009zt, Pan:2007nw, Boyle:2009dg} and TaylorT4 implementations \cite{Buonanno:2009zt} for the frequency and phase of the 3.5PN waveform \cite{Arun:2004hn}, as provided by the \verb|lalsuite| package \cite{lalsuite, swiglal}.

\begin{figure}
\centering
\includegraphics[width=0.95\columnwidth]{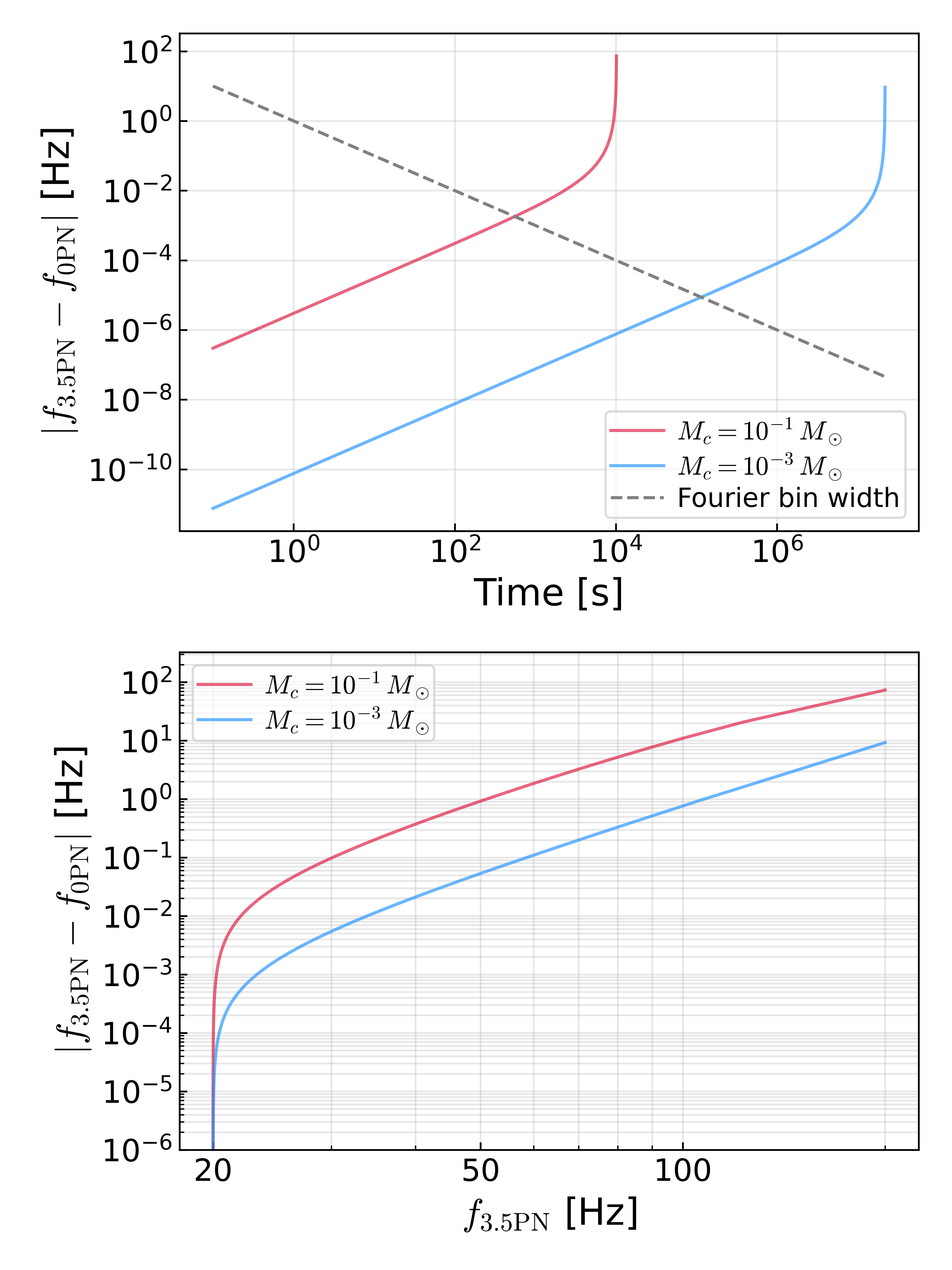}
\caption{
    Difference between the 0PN and 3.5PN waveform models for quasi-circular, non-spinning, equal-mass systems with $M_c = 10^{-1}M_\odot$ and $10^{-3}M_\odot$, and $f_0 = 20$~Hz. The upper (lower) panel shows the difference between the instantaneous frequencies as a function of time after a reference time (the instantaneous frequency of the 3.5PN model). 
}
\label{fig:PN_error}
\end{figure}

However, we choose to use 3.5PN waveforms rather than lower-order waveforms because, while the instantaneous frequency differences are often small, they accumulate over many waveform cycles and can produce large phase differences over a full waveform duration. Fig.~\ref{fig:PN_error} shows the difference between 3.5PN and 0PN waveforms with $f_0=20$~Hz and $M_c = 10^{-1}M_\odot$ and $10^{-3}M_\odot$. The upper panel demonstrates that the difference in instantaneous frequency between 3.5PN and 0PN waveforms grows rapidly in time. The grey line indicates the Fourier-bin width on the vertical axis, calculated as the reciprocal of the time shown on the horizontal axis. The intersection between the frequency residual curves and the grey line marks an approximate time after which a lower-order phase model would lose most of the signal power. It takes $\sim10^3$~s ($10^5$~s) for the frequency difference to exceed $10^{-3}$~Hz ($10^{-5}$~Hz) for the systems with $M_c=10^{-1}M_\odot$ ($10^{-3}M_\odot$). The lower panel shows the frequency difference between 3.5PN and 0PN waveforms as a function of the 3.5PN frequency. The 3.5PN frequency is approximately $21$~Hz ($20.1$~Hz) when the frequency difference reaches $10^{-3}$~Hz ($10^{-5}$~Hz) for systems with $M_c=10^{-1}M_\odot$ ($10^{-3}M_\odot$). In other words, after the signal has evolved by only $1$~Hz ($0.1$~Hz), the systematic error between 0PN and 3.5PN models is already sufficiently large to cause signal power leakage into adjacent Fourier bins. Other lower-PN-order models similarly introduce systematic error without reducing the number of free parameters or substantially decreasing the computational cost of generating waveforms with the TaylorF2 implementation used. 

When modeling the intrinsic strain amplitude in the waveform, we only consider the leading-order 0PN term, as is commonly adopted in searches for subsolar-mass binary BH signals \cite{LIGOScientific:2026wxz, LVK:2022ydq, Kacanja:2026byy}. The intrinsic strain amplitude evolution is then: 
\begin{align}
    h_0(t) &= \frac{4}{d}
    \left( \frac{G M_c}{c^2} \right)^{5/3}
    \left( \frac{\pi f(t)}{c} \right)^{2/3}
    \label{eq:h0} \\
    &\simeq
    \begin{aligned}[t]
    &1.6 \times 10^{-24}
    \left( \frac{d}{10\,\text{kpc}} \right)^{-1} \\
    & \times
    \left( \frac{M_c}{10^{-3} M_\odot} \right)^{5/3}
    \left( \frac{f(t)}{50\,\text{Hz}} \right)^{2/3} \, ,
    \end{aligned}
    \label{eq:h0_approx}
\end{align}
where $d$ is the distance to the binary system. 

The strain measured by a detector is the antenna-pattern-weighted sum of the two gravitational-wave polarizations,
\begin{equation}
    h(t) = F_+(t;\alpha,\delta,\psi) h_+(t) 
    + F_\times(t;\alpha,\delta,\psi) h_\times(t) ,
    \label{eq:detector_response}
\end{equation}
where $F_+$ and $F_\times$ are the detector antenna pattern functions, which depend on the source sky location $(\alpha,\delta)$, and the polarization angle $\psi$. For the quadrupole harmonic, we have
\begin{align}
    h_+(t) &= h_0(t) \frac{1+\cos^2\iota}{2} \cos\phi(t), \\
    h_\times(t) &= h_0(t) \cos\iota \sin\phi(t),
\end{align}
where $\iota$ is the inclination of the binary orbital angular momentum relative to the line of sight, and $\phi$ is the gravitational-wave phase.

\section{Methods}
\label{sec:methods}
A semicoherent search approach consists of two stages.  First, the full observing time $T_\mathrm{obs}$ is divided into segments of duration $T_{\rm coh}$. Each segment is coherently searched over using 0PN templates as described in Sec.~\ref{sec:coherent_analysis}. This coherent analysis uses a novel resampling implementation based on a non-uniform fast Fourier transform (NUFFT) as detailed in Sec. \ref{sec:resampling}. The coherent segments are sufficiently short that higher-order templates are not required for accurate phase integration.
The second stage of the analysis is a semicoherent combination of segments along 3.5PN frequency tracks as described in Sec.~\ref{sec:semicoherent_combination}. The template bank for this semicoherent stage is described in Sec.~\ref{sec:semicoherent_bank}.


\subsection{Coherent analysis over $T_{\rm coh}$}
\label{sec:coherent_analysis}
For each segment, we search for signals using a 0PN template which is sufficiently accurate to model the 3.5PN waveform model over the short duration of each segment. We use a 0PN model instead of a low-order Taylor expansion (as is common in other continuous-wave analyses) because it better approximates the non-linear chirping structure of the 3.5PN waveform. This better accuracy therefore permits longer coherent segments while retaining a low-dimensional model. At 0PN, the frequency evolution of gravitational waves emitted by a PBH binary during its inspiral is given by:
\begin{equation}
    f_{\rm 0PN}(t ;f_0, M_c, t_0) = f_0 \left[ 1-\frac{8}{3}\beta_0 (t-t_0) \right]^{-3/8} \, , \label{eq:f}
\end{equation}
where $\beta_0 = \beta(f_0, M_c)$ is as defined in Eq.~\eqref{eq:beta}. The corresponding phase evolution of this signal is:
\begin{equation}
    \phi_{\rm 0PN}(t;f_0, M_c, t_0) = -\frac{6\pi}{5} f_0 \frac{\left[1-\frac{8}{3}\beta_0(t-t_0)\right]^{5/8}}{\beta_0} + \phi_0 \, , \label{eq:phi}
\end{equation}
where $\phi_0$ is an arbitrary phase.  

We select a coherent segment length such that the mismatch between the 0PN model and the full 3.5PN signal is $\mathcal{M}_{\rm PN} \leq 0.05$. We define the usual noise-weighted inner product between two waveforms $a$ and $b$ as:
\begin{equation}
(a|b)
=
4 \, {\rm Re}
\int
\frac{a^{*}(f)b(f)}{S_n(f)}\,df \, ,
\end{equation}
where $*$ is the complex conjugate, and $S_n(f)$ is the one-sided noise spectral density. 
The overlap between the 3.5PN and 0PN waveforms is
\begin{equation}
\mathcal{O}(T_{\rm coh}, f_0, M_c)
=
\max
\frac{
\left(h_{\rm 3.5PN} \middle| h_{\rm 0PN}\right)
}{
\sqrt{
\left(h_{\rm 3.5PN} \middle| h_{\rm 3.5PN}\right)
\left(h_{\rm 0PN} \middle| h_{\rm 0PN}\right)
}
} ,
\end{equation}
where $h_{\rm 3.5PN}$ is a 3.5PN waveform, $h_{\rm 0PN}$ is a 0PN waveform, and the maximization is over relative time and phase shifts. The mismatch is defined as:
\begin{equation}
\mathcal{M}_{\rm PN}(T_{\rm coh}, f_0, M_c)
=
1 - \mathcal{O}(T_{\rm coh}, f_0, M_c).
\end{equation}
We require
\begin{equation}
\mathcal{M}_{\rm max} \equiv
\max_{f_0,M_c}
\mathcal{M}_{\rm PN}(T_{\rm coh}, f_0, M_c)
= 0.05
\label{eq:coherent_mismatch_max}
\end{equation}
for the parameter space considered, and it enforces a maximum coherence time of $60$~s; we choose $T_{\rm coh}=30~{\rm s}$ in this work as a conservative choice. It is also possible to choose different coherence times when analyzing different regions of the parameter space; we leave the detailed configuration choice studies for future work. 

When searching for signals, we need to search over $\beta$, which is not known a priori. Thus, we build a template bank of $\beta$ over the range $3\times10^{-8} \lesssim \beta  \lesssim 7\times10^{-4}$. We construct the template bank such that the mismatch between the $h_{\rm 3.5PN}$ and the closest $h_{\rm 0PN}$ template is less than the mismatch criterion we set earlier $\mathcal{M}_{\rm max} = \ 0.05$. We compute this mismatch numerically and uniformly place $\beta$ grids such that the mismatch condition is satisfied for the entire parameter space. We require 69 templates over $\beta$ to cover the range considered. 

\subsection{Resampling with non-uniform fast Fourier transforms}
\label{sec:resampling}

Within each coherent segment, we search for signals using the 0PN templates with resampling. Resampling is a signal-processing technique used to remove the expected phase evolution of a signal by changing the time coordinate in which the data is sampled \cite{tafti2006}. The signal that matches the template will then become monochromatic and the matched-filtering analysis can be performed using a Fast Fourier Transform (FFT) operation. Resampling was originally proposed for gravitational-wave signals in \cite{Smith:1987hh} and specifically for continuous gravitational waves in \cite{livasBroadbandSearchTechniques1989, Jaranowski:1998qm, Astone:2010ct, Braccini:2011zz}. More recent implementations are described in \cite{Astone:2014dea, Mastrogiovanni:2017xjr, Meadors:2017pefa, Singhal:2019dfn}, and used in observational searches, e.g., in \cite{LIGOScientific:2026qsb, LIGOScientific:2017ytx}. 

\begin{figure}
\centering
\includegraphics[width=0.95\columnwidth]{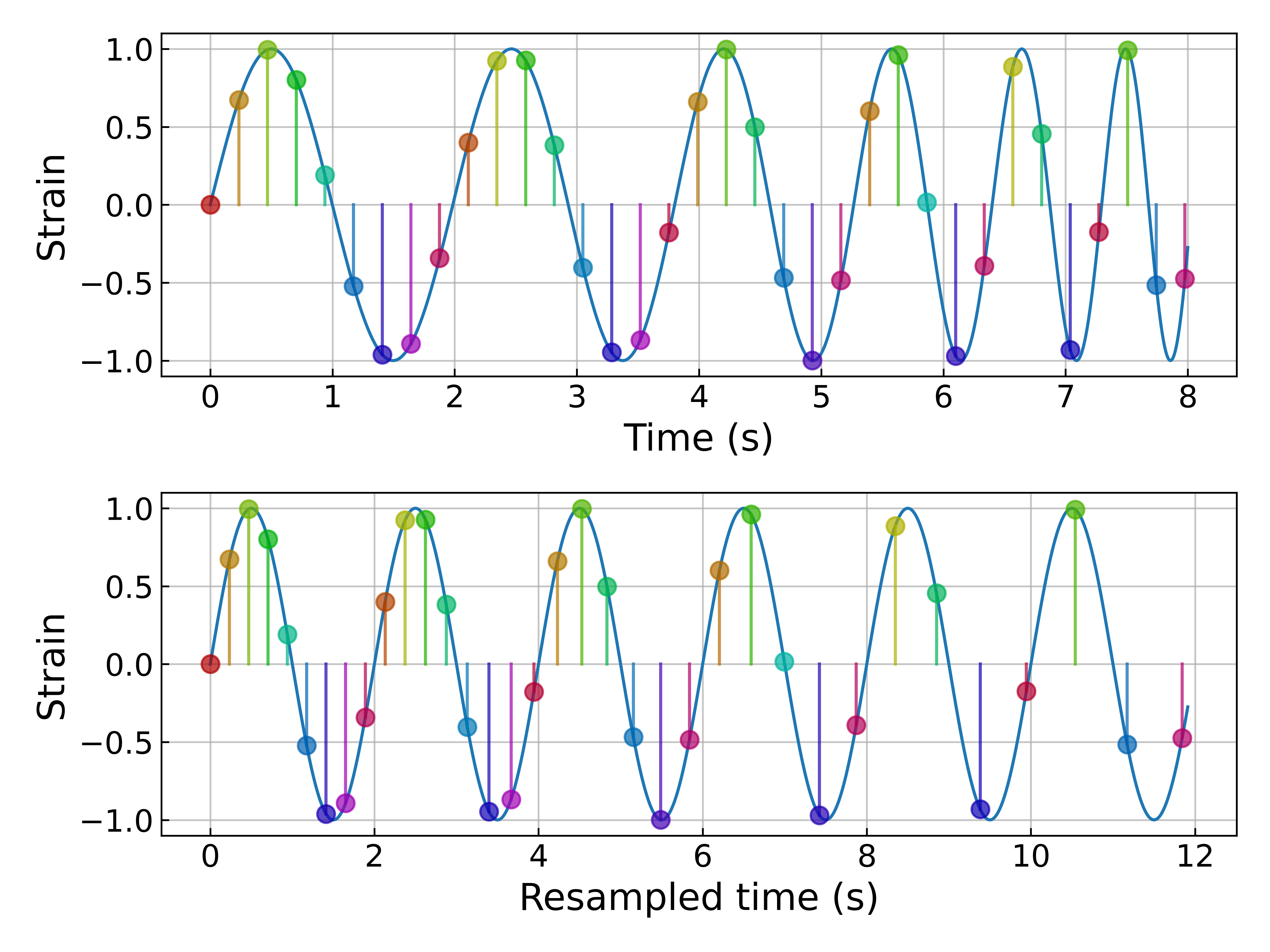}
\caption{
Resampling transforms a chirping, uniformly sampled signal (upper panel) into a monochromatic, non-uniformly sampled time series (lower panel). The colors correspond to the signal phase, with the same color indicating the same data point before and after the signal is resampled.
}
\label{fig:resampling_picture}
\end{figure}

\begin{figure}
\centering
\includegraphics[width=0.95\columnwidth]{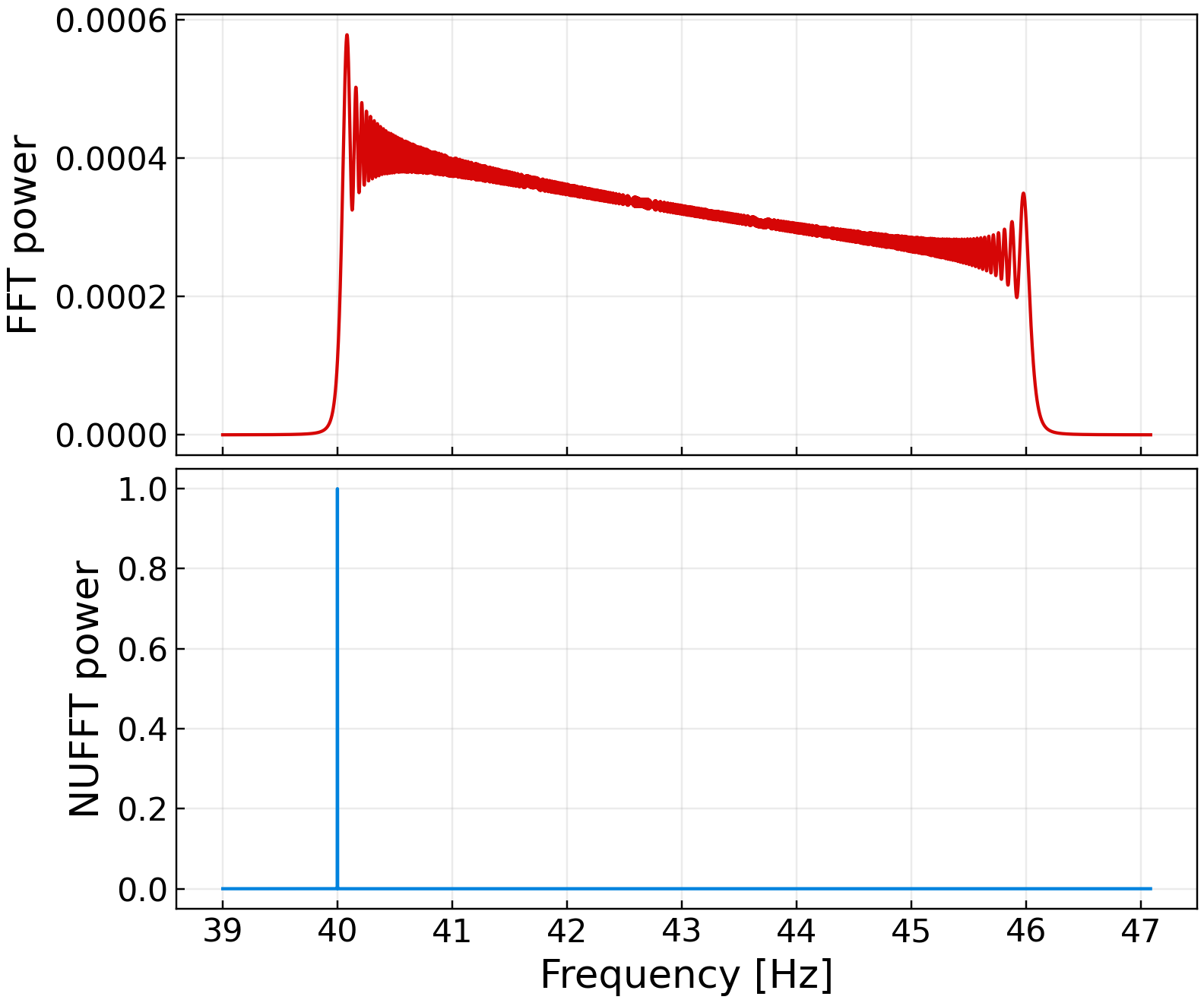}
\caption{
Resampling transforms a chirping signal with an evolving frequency (upper panel) to a monochromatic signal with power confined within a single Fourier bin (lower panel). 
}
\label{fig:resampling_spectra}
\end{figure}

The basic idea of resampling is to define a new time variable with respect to which the signal is monochromatic. For PBH binaries, we define the new time variable:
\begin{equation}
\tau = -\frac{3}{5} \frac{\left[1-\frac{8}{3}\beta_0 (t-t_0)\right]^{5/8}}{\beta_0} . \label{eq:new_t}
\end{equation}
Substituting Eq.~\eqref{eq:new_t} into Eq.~\eqref{eq:phi} gives:
\begin{equation}
\phi_{\rm 0PN}(t;f_0, M_c, t_0) = 2\pi f_0 \tau + \phi_0,\label{eq:monochromatic}
\end{equation}
i.e., the signal is monochromatic with respect to $\tau$. A demonstration of the basic principle of resampling is shown in Fig.~\ref{fig:resampling_picture}, where a signal with an evolving frequency is resampled and becomes monochromatic with respect to the resampled time coordinate. The Fourier-domain amplitude in each frequency bin is proportional to the matched-filtering output using the template with the corresponding value of $f_0$. A single FFT therefore simultaneously evaluates a bank of templates with $f_0$ spanning the bandwidth of the FFT. Fig.~\ref{fig:resampling_spectra} shows a signal with an evolving frequency such that the signal power is spread over many Fourier bins. After resampling, the signal power is concentrated in a single Fourier bin. 

However, after resampling, the signal is not uniformly sampled with respect to the new time variable $\tau$ and its FFT therefore cannot be computed directly. Different implementations effectively interpolate the signal onto a uniform grid in resampled time before applying the FFT. Previous implementations have used ``stroboscopic'' resampling \cite{Jaranowski:1998qm, Astone:2010ct, Singhal:2019dfn}, in which the original time series is sampled at a higher rate and nearest-neighbour interpolation is used to construct a uniformly sampled time series in $\tau$. However, nearest-neighbour interpolation is equivalent to convolution with a rectangular kernel and can introduce aliasing and errors in the resampled FFT. To mitigate these effects, the original time series must be oversampled and the resampled series subsequently downsampled when computing the FFT. This oversampling requirement increases the computational cost of stroboscopic resampling. We briefly explain how different resampling methods correspond to different choices of convolution kernel in Appendix~\ref{sec:resampling_convolution}.

Computationally efficient implementations of resampling use different kernel functions to interpolate the resampled time series onto a uniform time grid. Since we are ultimately interested in the Fourier transform of the resampled time series rather than the time series itself, we can draw on a relevant body of work in applied mathematics on computationally efficient choices of kernels. The operation of interpolating and Fourier transforming is known as a type-I NUFFT \cite{duttFastFourierTransforms1993, barnett2019}. Numerous NUFFT methods and implementations exist, employing different kernel functions. For early review articles with accessible introductions to the subject, see \cite{greengardAcceleratingNonuniformFast2004, wareFastApproximateFourier1998}. For modern implementations on CPUs and GPUs, see \cite{finufftFlatironinstituteFinufftNonuniform, shihCuFINUFFTLoadbalancedGPU2021, linJyhmiinlinPynufft2023, vaillantGhisvailPyNFFT2023, vanderplasJakevdpNfft2023, ouGNUFFTWAutoTuningHighPerformance2017} and the references within. Throughout this work, we use the fiNUFFT implementation described in Ref.~\cite{barnett2019}. 

\begin{figure}
\centering
\includegraphics[width=0.95\columnwidth]{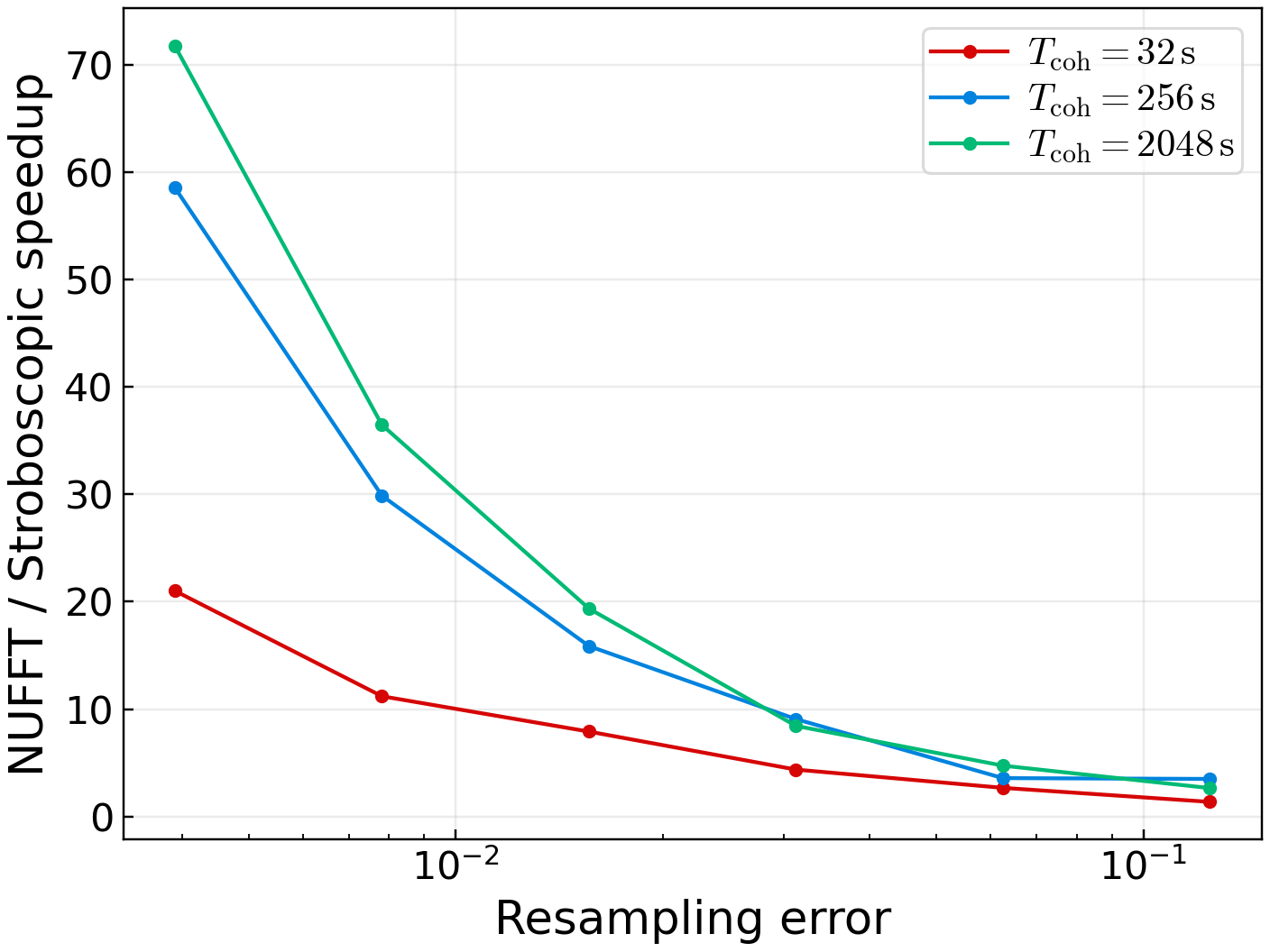}
\caption{
Speedup of using NUFFT relative to stroboscopic resampling as a function of resampling error, for $f_0 = 20\,\mathrm{Hz}$, $M_c = 10^{-2} M_\odot$, $T_\text{coh} = 32$~s, 256~s, 2048~s, and sampling rate $f_\mathrm{samp} = 512 \, \mathrm{Hz}$ for the resampled time series. 
}
\label{fig:NUFFT_speedup}
\end{figure}

The speedup of NUFFT relative to stroboscopic resampling as a function of the desired resampling accuracy is benchmarked using four threads on a commercial Intel 255U CPU, with the results shown in Fig. \ref{fig:NUFFT_speedup}. Note, here the resampling error refers to the worst case asymptotic bound for a particular algorithm rather than a definitive estimate of the error in this computation. Significant speedups can be achieved by using NUFFT implementations. Some stroboscopic resampling implementations in continuous-wave searches use parameters corresponding to an accuracy of $\approx 0.025$, for which our benchmark shows a speedup of $\mathcal{O}(10)$. This estimate should be regarded as indicative, since the actual speedup depends on the coherence length, CPU model, and numerical implementations. 

\subsection{Semicoherent combination over $T_{\rm obs}$}
\label{sec:semicoherent_combination}
After the coherent analysis in each segment using resampling, the data are represented by a collection of resampled Fourier amplitudes $\tilde{x}_{ijk}$, where the indices correspond to the $i$-th segment, $j$-th Fourier bin, and $k$-th value of $\beta$. The Fourier amplitudes along the frequency track of a candidate signal can then be combined using any suitable semicoherent method, including peakmaps \cite{Astone:2014esa, Miller:2018rbg, Piccinni:2018akm}, Hough-transform \cite{Krishnan:2004sv, Astone:2014esa, Mendell_stackslide_hough, Miller:2018rbg}, $\mathcal{F}$-statistic \cite{Jaranowski:1998qm}, or Viterbi-based approaches \cite{Suvorova:2016rdc, Sun:2017zge, Alestas:2024ubs}. Note that these methods would need to be modified to operate on higher-dimensional tracks since our spectra are three-dimensional rather than the usual two-dimensional time-frequency maps. In this work, we describe a noise-weighted stack-slide implementation \cite{Brady:1998nj, Mendell_stackslide_hough}. 

The stack-slide method accumulates the power along a candidate frequency evolution track. The parameters of the semicoherent frequency template are $\boldtheta=\{M_c, t_0\}$, where we set $f_0=40$~Hz to the lower boundary of the analyzed frequency band. In the $i$-th segment, this model corresponds to a frequency of $f_j(\boldtheta)$ and a 0PN resampling template $\beta_k(\boldtheta)$ used for the coherent analysis within the segment (we will refer to these as $f_j$ and $\beta_k$). This frequency evolution track is spread across $N_{\rm seg}(\boldtheta)$ segments, and the segments, Fourier bins and $\beta$ templates of the track are denoted by $(i,j,k) \in C(\boldtheta)$. 

For a given segment, we define the normalized power as:
\begin{equation}
P_{ijk}(\boldtheta)
=
\frac{4\left|\widetilde{x}_{ijk}\right|^2}
{T_{\rm coh} S_{ijk}^{\rm eff}},
\label{eq:normalized_power}
\end{equation}
where $S_{ijk}^{\rm eff}$ is the effective one-sided noise power-spectral density (PSD) in the $i$-th segment and $j$-th Fourier bin after the noise has also been resampled with the $k$-th $\beta$ template. 
The relative weight of segment $i$ is:
\begin{equation}
w_{ijk}(\boldtheta)
\propto
\frac{A_i^2(\boldtheta)\Gamma_i}
{S_{ijk}^{\rm eff}}, \label{eq:weights}
\end{equation}
where $A_i^2(\boldtheta)$ is the squared signal amplitude averaged over the segment and $\Gamma_i$ is a correction factor accounting for the reduction in signal power caused by the window function. The weights are normalized for each frequency track to satisfy: 
\begin{gather}
    \sum_{{(i,j,k)}\in C(\boldtheta)}w_{ijk}^2(\boldtheta)=1 .
\end{gather}
Weighting different segments is crucial for PBH signals as they chirp across large frequency ranges such that the signal amplitude and PSD can vary significantly. The short coherent times mean that the search is generally not sensitive to sky location, and in this case $A_i^2(\boldtheta)$ depends only on the intrinsic signal parameters. However, it is also possible to target the search to a particular sky location and account for a particular detector antenna response function $F_{+/\times}(t;\alpha,\delta,\psi)$ in the weighting function, leading to improved sensitivity (discussed later in this section).

The semicoherent statistic is then:
\begin{equation}
n_\sigma(\boldtheta)
=
\sum_{{(i,j,k)}\in C(\boldtheta)}
w_{ijk}(\boldtheta)
\frac{P_{ijk}-\mu_i}{\sigma_i}
,
\label{eq:semicoherent_statistic}
\end{equation}
where $\mu_i$ and $\sigma_i$ are the mean and standard deviation of $P_i$ for the background. For stationary Gaussian noise, $P_{ijk}$ is distributed according to a $\chi_2^2$ distribution and $\mu_i=2$ and $\sigma_i=2$, but for detector noise they can be computed empirically. 

\begin{figure}
\centering
\includegraphics[width=0.95\columnwidth]{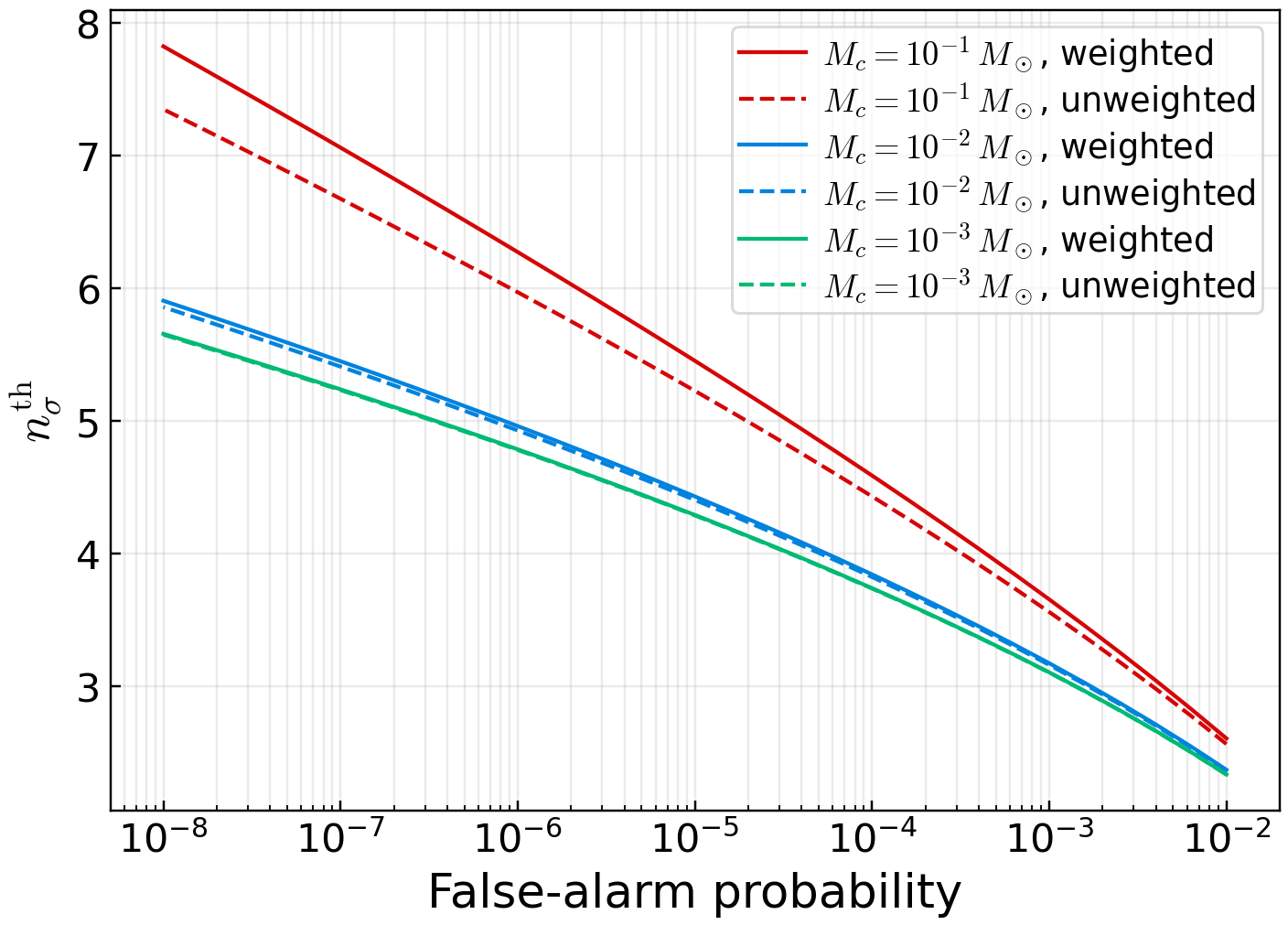}
\caption{
Threshold $n_\sigma^{\rm th}$ as a function of false-alarm probability for the weighted and unweighted distributions. Results are obtained for signals with an initial frequency of $f_0=40$~Hz and chirp masses of $M_c = 10^{-1}M_\odot, 10^{-2}M_\odot$, and $10^{-3}M_\odot$. 
}
\label{fig:chi2_threshold}
\end{figure}

Under the null hypothesis $H_0$, $n_\sigma$ has zero mean and unit variance. For equal weights, its distribution is a scaled $\chi_{2N_{\rm seg}}^2$ distribution. For unequal weights it is a generalized chi-squared distribution, a weighted sum of $N_{\rm seg}$ different $\chi_2^2$ distributions, and we evaluate its distribution and survival function numerically. Given a false-alarm probability ($p_{\rm fa}$), we define the detection threshold $n_\sigma^{\rm th}$ such that:
\begin{gather}
    p(n_\sigma > n_\sigma^{\rm th} | H_0) = p_{\rm fa} . \label{eq:n_sigma^th}
\end{gather}
The weighting does not significantly affect the $n_\sigma^{\rm th}$ for a given $p_{\rm fa}$. This is shown in Fig.~\ref{fig:chi2_threshold} which compares $n_\sigma^{\rm th}$ for equal and unequal weights for signals with an initial frequency of $f_0=40$~Hz and chirp masses of $M_c = 10^{-1}M_\odot, 10^{-2}M_\odot$ and $10^{-3}M_\odot$. The weighted $n_\sigma^{\rm th}$ is larger by $5\%, 0.7\%, 0.1\%$ for $M_c = 10^{-1}M_\odot, 10^{-2}M_\odot, 10^{-3}M_\odot$, respectively. For the majority of the parameter space the $n_\sigma^{\rm th}$ computed assuming uniform weights is therefore a close approximation of the weighted threshold. 

The weighting can, however, increase the recovered $n_\sigma$ from a given signal. In the presence of a signal 
$P_{ijk}$ is distributed according to a non-central $\chi_2^2$ distribution with $\mu_i=2+\lambda_i$, where $\lambda_i$ is the non-centrality parameter and is equal to the normalized power [defined in Eq.~\eqref{eq:normalized_power}] of the signal in a particular segment. In this case, $n_\sigma$ is distributed according to a generalized chi-squared distribution, a weighted sum of $N_{\rm seg}$ number of non-central $\chi_{2}^2$ distributions, and has a total non-centrality $\Lambda(\boldtheta) = \sum_{(i,j,k) \in C(\boldtheta)} \lambda_i$.

\begin{figure}
\centering
\includegraphics[width=0.95\columnwidth]{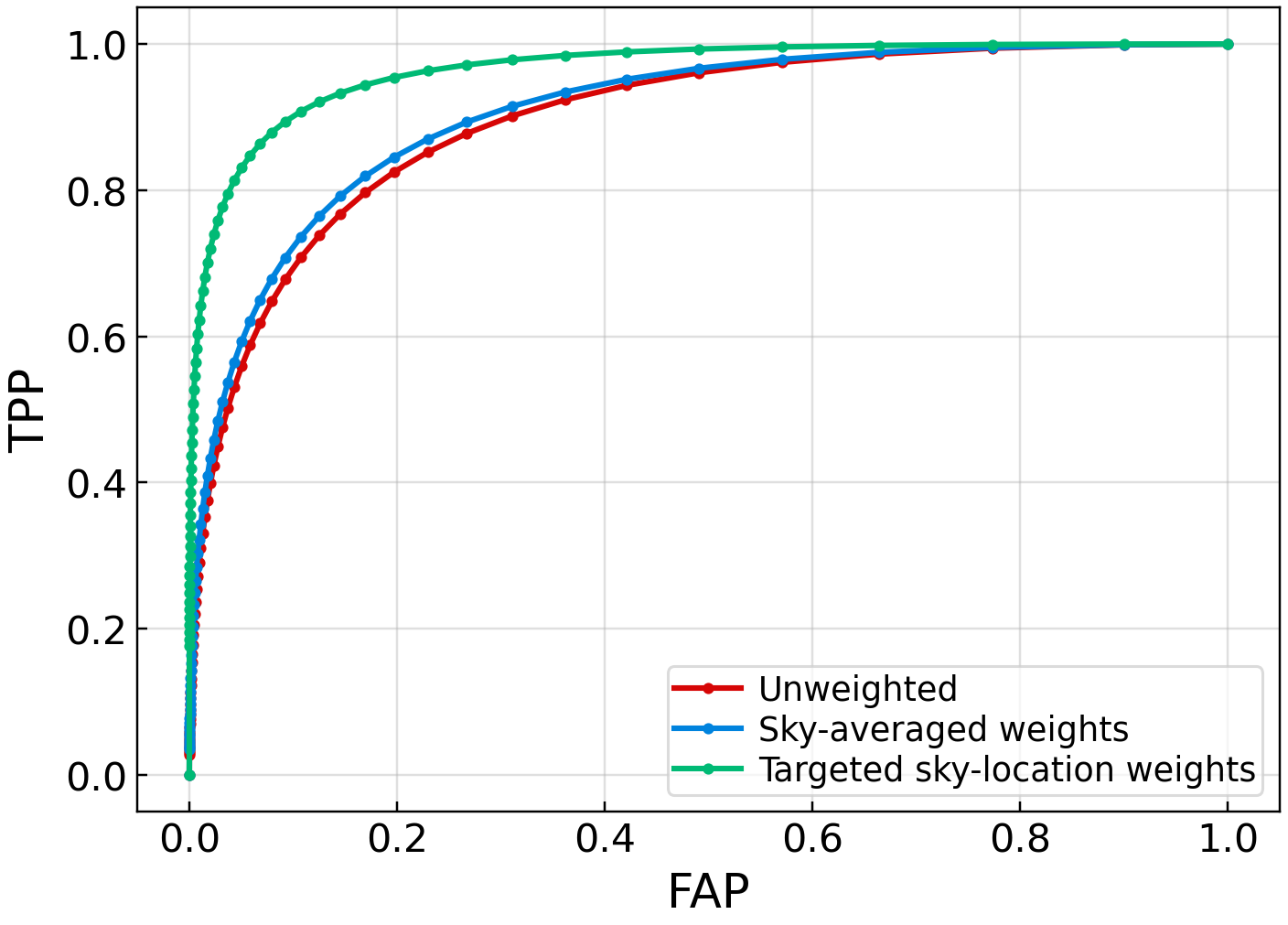}
\caption{
ROC curve for signals with $M_c = 10^{-3}M_\odot$, $f_0=40$~Hz, $f_{\rm high}=60$~Hz with randomized sky locations and $d=50$~kpc. The curves demonstrates that accounting for the signal weights improves detectability, especially when targeting a specific sky location. 
}
\label{fig:roc}
\end{figure}

The improvement in overall signal detectability from weighting is shown using a receiver operating characteristic (ROC) curve in Fig.~\ref{fig:roc}, which plots the detection probability $p_{\rm det}$ (also commonly referred to as the true positive probability) against the $p_{\rm fa}$.  Signals with randomized sky locations are injected at a fixed distance of $d=50$~kpc and recovered using the LIGO Livingston detector. The signals are recovered assuming idealized colored Gaussian noise at the Advanced LIGO design sensitivity \footnote{The amplitude spectral density is given by \texttt{aligo\_O4high.txt} \cite{LIGOT2000012v2NoiseCurves} and is representative of the sensitivity achieved during the fourth gravitational-wave observing run~\cite{Capote:2024rmo}.}, and the detection thresholds for different $p_{\rm fa}$ are determined from the theoretical noise-only distribution. We compute $p_{\rm det}$ as the fraction of injected signals that exceed $n_\sigma^{\rm th}$. A search where the antenna-responses are marginalized over but the intrinsic amplitude evolution is accounted for leads to a small improvement in sensitivity compared to the unweighted search. Targeting a particular sky location, however, markedly increases the detectability of the signal and demonstrates the improved sensitivity of targeted searches even though the short coherence lengths of the search allow for blind searches. 

Throughout the rest of this work, we define a signal as detected if it satisfies $p_{\rm fa}=10^{-6}$ and $p_{\rm det}=0.95$. We can numerically evaluate the generalized $\chi^2$ distribution with the weights $\{w_{ijk}\}$ to define the $\Lambda^{\rm det}_{N_{\rm seg}}$ as the non-centrality required for a signal to be detected:
\begin{equation}
    p\!\left(
        n_\sigma>n_\sigma^{\rm th}
        \mid \Lambda^{\rm det}_{N_{\rm seg}}=\sum_{(i,j,k)\in C(\boldtheta)} \lambda_i, \{w_{ijk}\}
    \right)
    =p_{\rm det}. 
    \label{eq:non-centrality_threshold}
\end{equation} 
The robustness of our generalized chi-squared distribution expectation is verified by comparing to numerical injections in Sec.~\ref{sec:injection}.

\subsection{Semicoherent template bank}
\label{sec:semicoherent_bank}

The semicoherent integration stage requires a template bank of 3.5PN waveforms distinct from the 0PN $\beta$ templates used in each coherent segment. We define the semicoherent mismatch as:
\begin{equation}
    \mathfrak{m}(\boldtheta, \boldtheta_s) = 1-\frac{n_\sigma(\boldtheta|\boldtheta_s)}{n_\sigma(\boldtheta_s|\boldtheta_s)} \, , \label{eq:semicoherent_mismatch}
\end{equation}
where $n_\sigma(\boldtheta|\boldtheta_s)$ is the noise-averaged semicoherent statistic for a template $\boldtheta$ given the presence of a signal with true parameters $\boldtheta_s$. We require:
\begin{equation}
    \mathfrak{m}_{\rm max} \equiv \max_{\boldtheta, \boldtheta_s}\mathfrak{m}(\boldtheta, \boldtheta_s) = 0.05 . \label{eq:semicoherent_mismatch_max}
\end{equation}

We then estimate the required size of the template bank using a geometric approach, commonly adopted in semicoherent searches \cite{Prix:2007ks,Wette:2014tca}. Other algorithms developed for
compact-binary template banks \cite{Brown:2012nn}, including
MANIFOLD-style, stochastic, or hybrid bank-generation strategies
\cite{Hanna:2022zpk,Hanna:2024tom,Kacanja:2024pjh,Roy:2017oul,Roy:2017qgg},
could be adapted to the semicoherent mismatch and applied. In the geometric approach, we define a metric:
\begin{equation}
g_{ab}(\boldtheta_s)
=
-\frac{1}{2n_\sigma(\boldtheta_s;\boldtheta_s)}
\left.
\frac{\partial^2 n_\sigma(\boldtheta;\boldtheta_s)}
{\partial\boldtheta^a\partial\boldtheta^b}
\right|_{\boldtheta=\boldtheta_s} ,
\label{eq:semicoherent_metric_tensor}
\end{equation}
and use it to approximate Eq. \eqref{eq:semicoherent_mismatch} for a small offset $\Delta\boldtheta$ from the true signal parameter $\boldtheta_s$: 
\begin{align}
\mathfrak{m}(\boldtheta_s+\Delta\boldtheta;\boldtheta_s)
&\simeq  g_{ab}(\boldtheta_s)\Delta\boldtheta_a\Delta\boldtheta_b,
\label{eq:semicoherent_metric_expansion}
\end{align}
where summation over the repeated indices $a$ and $b$ is implied. Here we numerically differentiate the TaylorF2 signal model to obtain the non-uniform metric. 
We use the metric to compute the number of templates required:
\begin{equation}
    N \simeq \Theta
    {\mathfrak{m}_{\max}}^{-n/2}
    \int_{\rm bank}
    \sqrt{\det g(\boldtheta)}
    \,dM_c\,dt_{\rm 0},
    \label{eq:standard_template_count}
\end{equation}
where $n$ is the dimension of the bank and $\Theta$ is the dimensionless
covering factor set by the lattice geometry.  In two dimensions,
$\Theta=1/2$ for a square covering and
$\Theta=2/(3\sqrt{3})$ for a hexagonal covering \cite{Jaranowski:2005hz}. For a hexagonal covering, this gives
an estimate of $2\times 10^{10}$ templates required. 

\section{Horizon distance}
\label{sec:distance_sens}

In this section, we compute the horizon distance of the semicoherent search at the Advanced LIGO design sensitivity \cite{LIGOT2000012v2NoiseCurves}, examine its dependence on the choice of analysis frequency band, and justify our choice of the 40--60 Hz band. We define the horizon distance as the maximum distance at which a signal is
detectable with $p_{\rm det}=95\%$ and $p_{\rm fa}=10^{-6}$. We first derive an expression for the horizon distance for a fully coherent analysis over the in-band signal duration. After averaging over sky position,
inclination, polarization, and Fourier-bin offset, the coherent (matched-filter)
sensitivity to distance is:
\begin{align}
  d_{\rm coh}(\boldtheta, f_{\rm high}) =& 
  \frac{0.007517 \, c \, \beta_0}
  {f_0^{8/3} \left(\Lambda_1^{\rm th}\right)^{1/2}} \nonumber\\
  \times& \left(\int_{t_0}^{t_{\rm end}}
  \frac{f_{\rm 3.5PN}(t, \boldtheta)^{4/3}}
  {S_n[f(t,\boldtheta)]}dt\right)^{1/2}
\end{align}
where $\Lambda_1^{\rm th}=46.5$ [computed by evaluating Eq.~\eqref{eq:non-centrality_threshold}], $f_{\rm 3.5 PN}(t, \boldtheta)$ is the 3.5PN TaylorF2 frequency track, and $t_{\rm end}$ is the end time of the frequency track, taken to be the minimum between the time when the signal reaches the upper limit of the analysis frequency band $f_{\rm high}$, and the entire observation time $T_{\rm obs}=1$~year. 

The horizon distance obtained using the semicoherent approach accounts for the mismatches within each coherent segment $\mathcal{M}_{\rm max}$, the mismatches from the semicoherent template bank $\mathfrak{m}_{\rm max}$ over $T_{\rm obs}$, and the loss in sensitivity due to the incoherent combination. The latter is quantified by $\mathcal{P}$, which is defined as:
\begin{equation}
    \mathcal{P}(f_0, f_{\rm high})
    \equiv
    \left(
        \frac{\Lambda_1^{\rm det}}
             {\Lambda_{N_{\rm seg}}^{\rm det}}
    \right)^{1/2},
    \label{eq:semicoherent_penalty}
\end{equation}
where $N_{\rm seg}$ depends on $f_0$ and $f_{\rm high}$. For equal weights, in the limit of $N_{\rm seg}\gg1$, the $\Lambda_{N_{\rm seg}}$ statistic approaches a Gaussian distribution and the required noncentrality scales as $\Lambda_{N_{\rm seg}}^{\rm th}\propto N_{\rm seg}^{1/2}$. In this limit, we recover the standard semicoherent scaling $\mathcal{P} \propto {N_{\rm seg}}^{-1/4}$, although throughout this work we calculate $\mathcal{P}$ directly from the generalized $\chi^2$ distribution. The horizon distance that can be obtained using the semicoherent approach is then:
\begin{equation}
\begin{aligned}
    d_{\rm semi}(\boldtheta, f_{\rm high})
    &= \mathcal{P}(f_0, f_{\rm high})(1-\mathcal{M}_{\rm max})(1-\mathfrak{m}_{\rm max})^{1/2} \\
    &\quad \times 
         d_{\rm coh}(\boldtheta,f_{\rm high}).
    \label{eq:semicoherent_dist_sens}
\end{aligned}
\end{equation}

\begin{figure}
    \centering
    \includegraphics[width=\columnwidth]{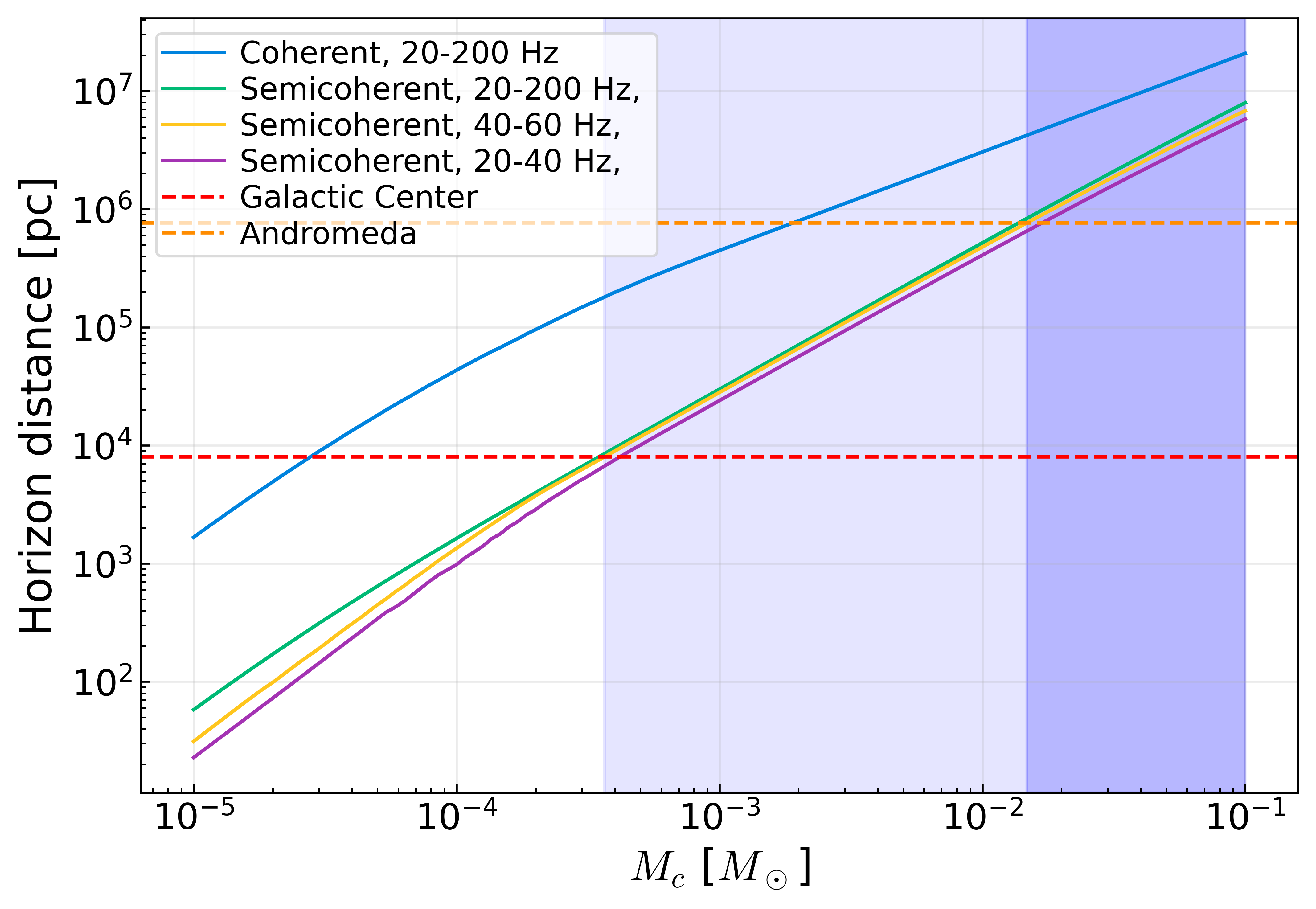}
    \caption{
        Horizon distance obtained using a semicoherent approach as a function of the system chirp mass $M_c$. The red (orange) dashed line corresponds to the distance to the Galactic Center (Andromeda). The shaded regions show the portion of the parameter space where the horizon distance extends beyond the Galactic Center (lighter) and Andromeda (darker), indicating the search could feasibly detect PBHs within these regions.
    }
    \label{fig:semicoherent_sensitivity_1d}
\end{figure}

A choice of analysis frequency band is defined by $f_{\rm low}$ and $f_{\rm high}$, between which we track a frequency evolution template. We compute the horizon distance by optimizing the initial frequency $f_0$ within the analysis band and show the results in Fig.~\ref{fig:semicoherent_sensitivity_1d}.\footnote{For a binary with the same chirp mass that remains in the analysis band for longer than the maximum observing time of one year, targeting a later stage of the inspiral improves the sensitivity.} The dashed lines indicate the distances to the Galactic Center and the Andromeda galaxy, two locations of particular interest for PBH searches because of their high dark-matter densities. This is because, if PBHs constitute a fraction of the dark matter, their abundance is generally expected to trace the underlying dark-matter distribution in these regions \cite{Green:2020jor, Carr:2020xqk}.

The blue curve, corresponding to a fully coherent analysis over a 20--200 Hz analysis band, shows the theoretical optimum but is computationally prohibitive. The horizon distances for three choices of semicoherent analysis band are shown, each of which has a different $T_{\rm coh}$ to ensure that the coherent mismatch condition [Eq.~\eqref{eq:coherent_mismatch_max}] is satisfied in the analysis band. The figure shows that a wide analysis band of 20--200 Hz yields the furthest horizon distance but also requires a higher computational cost to compute the spectra and store the data. The 40--60 Hz band is therefore a good compromise, achieving a horizon distance almost identical to that of the 20--200 Hz band for much of the parameter space while being slightly more sensitive than the 20--40 Hz band. The close proximity of all three choices is because the extra power gained by a larger analysis band is broadly offset by the shorter coherence times and greater number of segments. The figure motivates the chirp mass parameter range of $5 \times 10^{-4} M_\odot \lesssim M_c \leq 10^{-1} M_\odot$. The lower mass limit represents the lightest PBH systems the search is sensitive to within the Galactic Center (light shaded), and the upper mass limit represents the systems which are conventionally searched for by short-duration subsolar-mass searches \cite{LIGOScientific:2026wxz, Kacanja:2026byy}.

\section{Injection verification}
\label{sec:injection}

\begin{figure*}
    \centering
    \includegraphics[width=0.8\textwidth]{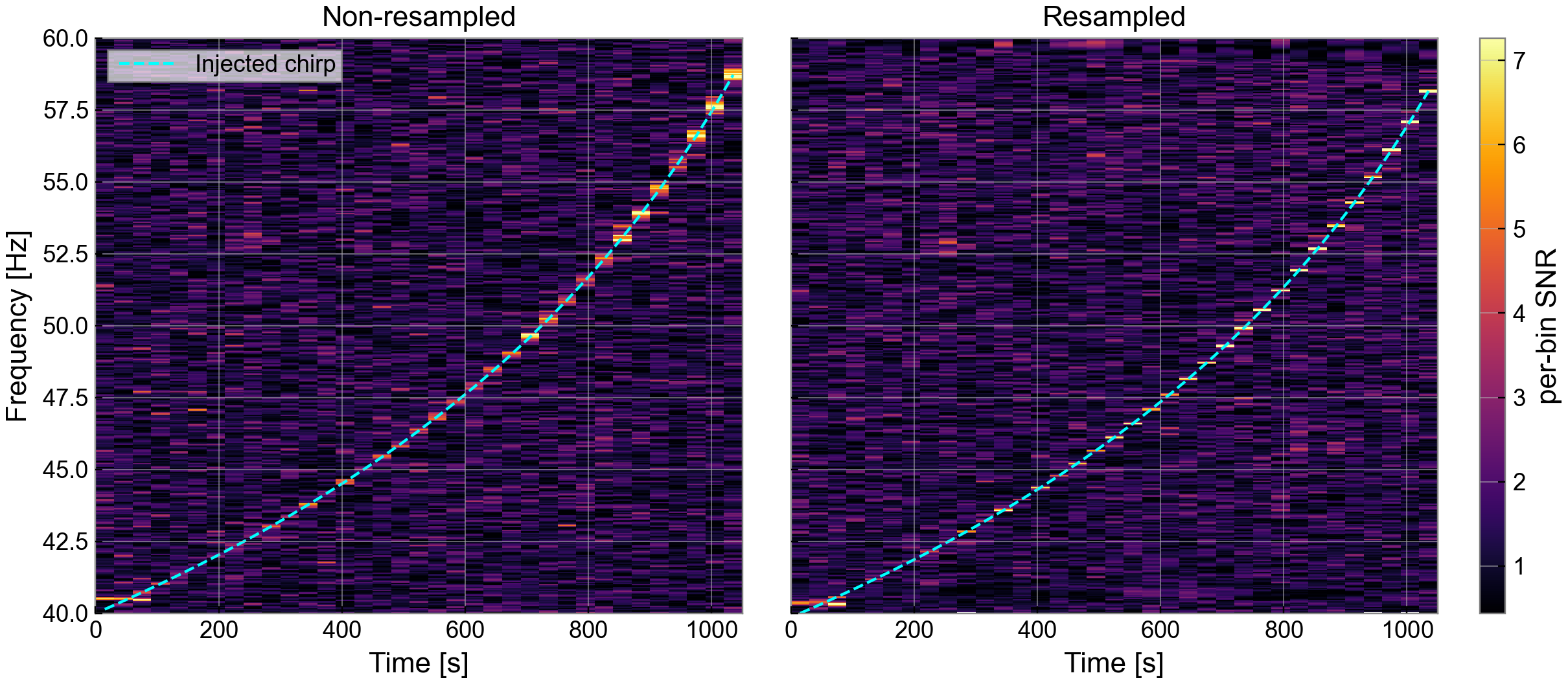}
    \caption{
        Frequency-time spectrogram of a 3.5PN inspiral signal with $f_0 = 40$~Hz and $M_c=10^{-1}M_\odot$, injected into detector noise from the LIGO Hanford detector collected during O4a. The left panel shows that the non-resampled signal spreads its power over multiple frequency bins (especially at later times) because of the frequency evolution of the signal, whereas the right panel shows that resampling is able to confine the majority of the signal power within individual frequency bins.
        }
    \label{fig:injection_spectra}
\end{figure*}

We validate the theoretical sensitivity projection by injecting simulated 3.5PN inspiral signals with parameters $f_0=40$~Hz and $M_c=10^{-1}M_\odot$. The signals are analyzed over the 40--60~Hz frequency band using 35 coherent segments of duration $T_{\rm coh}=30$~s, corresponding to a total signal duration of 1050~s.

We inject these signals into 100 independent stretches of detector noise, with randomized sky locations, polarization angles, and inclinations. The noise is drawn from the Hanford detector data collected during the first part of the fourth observing run (O4a) \cite{LIGOScientific:2025snk}, requiring that each 1050~s stretch remains in observing mode continuously, lies at least 10~s from any transient gravitational-wave signal identified in the GWTC-4 catalog \cite{LIGOScientific:2025slb}, and has not been used for any previous injections.

For each injection, we read the detector strain using \texttt{GWpy} \cite{Macleod:2021goi} and generate the injected waveform using the time-domain TaylorT4 approximant. We recover the signals using the procedure described in Sec.~\ref{sec:methods} but perform the coherent and semicoherent analyses directly at the injected parameter values. We do not demonstrate a fully agnostic search over the parameter space, which would require substantially greater computational cost, but instead validate the analysis procedure and its agreement with theoretical predictions.

Fig.~\ref{fig:injection_spectra} shows a time-frequency representation of an example injection. The resampling procedure confines the signal power to approximately a single frequency bin within each coherent segment. Without resampling, the signal shows a frequency evolution within each coherent segment, and a standard FFT distributes the signal power over multiple frequency bins.

\begin{figure}
    \centering
    \includegraphics[width=0.8\columnwidth]{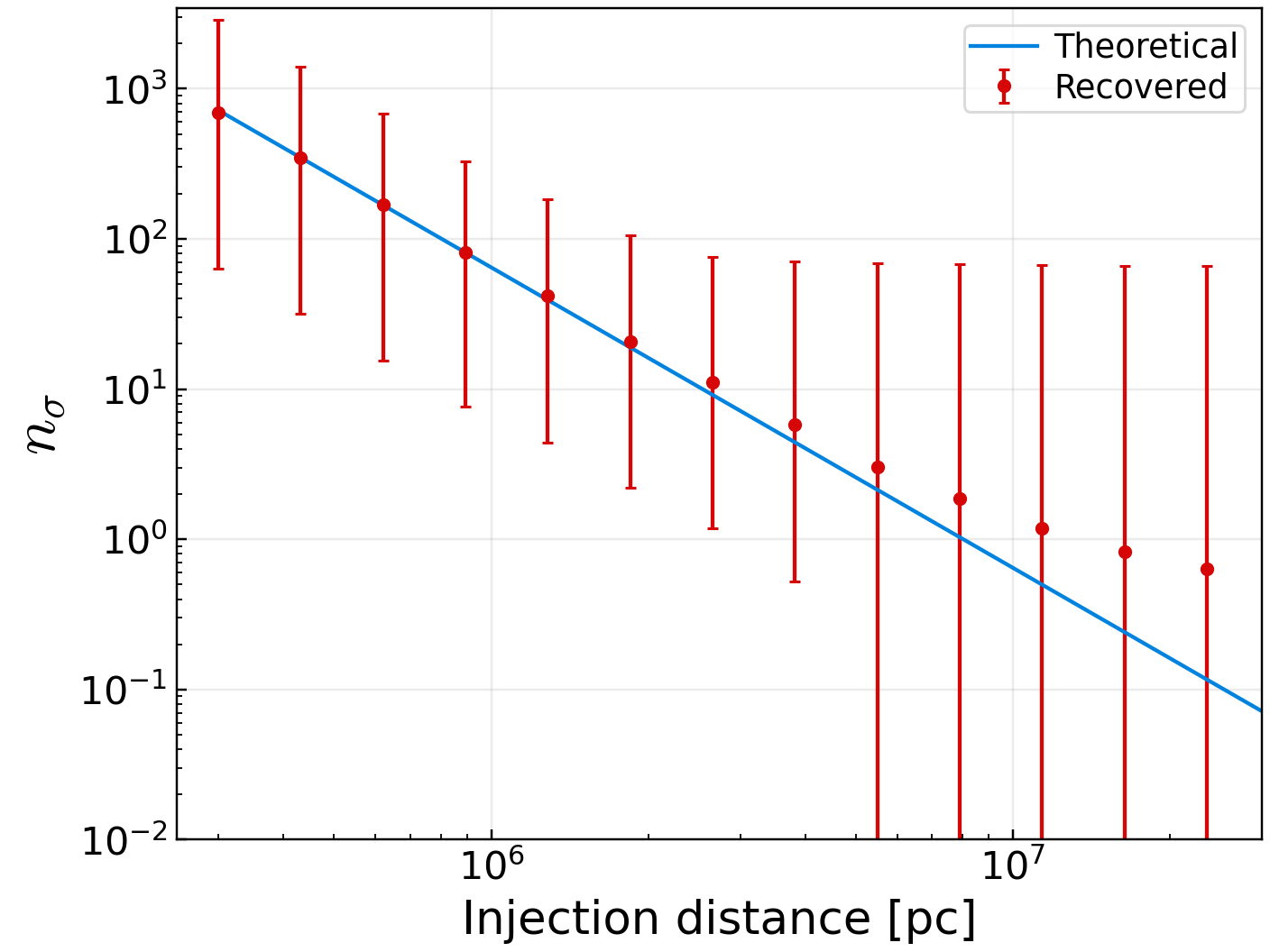}
    \caption{
        Semicoherent detection statistic $n_\sigma$ as a function of distance. The blue curve shows the theoretical prediction, while the red points with error bars indicate the medians and 90\% confidence intervals obtained from 100 injections. The signals are generated with $f_0=40$~Hz and $M_c=10^{-1}M_\odot$, with randomized sky locations and polarization angles, and injected into O4a Hanford detector noise. The results agree well with the theoretical prediction until the noise dominates the signal, at which point non-Gaussian noise fluctuations introduce a systematic bias in the recovered detection statistic.
    }
    \label{fig:detector_noise_injections}
\end{figure}

The results of the 100 injections at each of 13 different distances are shown in Fig.~\ref{fig:detector_noise_injections}, where the red points with error bars indicate the medians and 90\% confidence intervals. In the signal-dominated regime, where the injected signal lies above the noise floor, the recovered statistic agrees well with the theoretical prediction (blue curve). At distances $\gtrsim 6\times10^6$~pc, however, the recovered values are systematically larger than the theoretical prediction, as the statistic becomes increasingly dominated by detector noise and affected by low-power glitches or other non-Gaussian transients. This is also reflected in the increasing sizes of the error bars, whose lower bounds extend to negative values of $n_\sigma$ and therefore are not well represented on the logarithmic axis. Overall, these injections provide a proof of concept for the analysis procedure and demonstrate that the recovered semicoherent statistic exhibits the expected scaling with source distance.

\section{Computational cost}
\label{sec:cost}
The computational cost of the search is dominated by two stages: constructing the resampled spectra and evaluating the semicoherent stack-slide statistic. For each value of the coherent resampling parameter $\beta$, we compute NUFFTs over $N_{\rm seg} = \lceil T_{\rm obs}/T_{\rm coh}\rceil$ coherent segments, where $\lceil \cdot \rceil$ denotes the ceiling function. Each coherent segment contains $\lceil T_{\rm coh} \cdot f_{\rm samp}\rceil$ samples, where $f_{\rm samp}$ is the sampling rate of the time series. If we fix the desired NUFFT accuracy, the computational cost scales as:
\begin{equation}
C_{\rm NUFFT}
\propto
N_\beta T_{\rm obs} f_{\rm samp}
\log\left(T_{\rm coh} f_{\rm samp}\right) ,
\end{equation}
here, $f_{\rm samp}$ is the sampling rate of the resampled time series, taken here to be 512 Hz, and $N_\beta=69$ is the number of beta templates for our search (calculated in Sec.~\ref{sec:coherent_analysis}).
In practice, we can benchmark this directly. On a commercial Intel Core Ultra 7 255U CPU utilizing four threads, the current implementation is able to analyze $\mathcal{O}(10^6)$ 30-s segments per hour and thus would compute all the NUFFT spectra in $\mathcal{O}(10^2)$ hours. This is parallelizable and we could expect speedups of one to two orders of magnitude for a GPU implementation.

After the resampled spectra are generated, evaluating the stack-slide statistic is computationally simple and consists of reading powers from the resampled spectra and accumulating weighted sums. It is therefore expected to be limited by memory bandwidth. The search would require evaluation times of $\mathcal{O}(1)\,\mathrm{hr}$ on a commercial CPU. For more details of this estimate see Appendix \ref{sec:stackslide_cost}. 

A practical limitation is memory size: storing the resampled spectra for a year of observing data from one detector would require $\approx 200$~GB. This exceeds the memory available on a single high-end GPU. A device with a memory size of 80 GB could store the spectra from $\approx 165$ days of observing data. Templates with durations longer than this could be handled by caching partial sums, which adds only a small memory overhead compared to storing the spectra themselves.


\section{Conclusion}
\label{sec:conclusion}
In this work, we present a novel implementation of resampling based on type-I NUFFTs. The method evaluates the Fourier spectrum directly from samples that are non-uniform in the transformed time coordinate, thereby combining the resampling and Fourier-transform operations without constructing a uniformly
sampled intermediate time series. It avoids the nearest-neighbour interpolation and oversampling required by stroboscopic resampling. At numerical accuracies representative of existing stroboscopic implementations, our benchmarks show a speedup of $\mathcal{O}(10)$--$\mathcal{O}(10^2)$. The precise gain depends on the coherent-segment duration, computing hardware, and requested numerical tolerance.

To demonstrate the applicability of this technique, we have constructed a frequentist semicoherent search for long-duration inspirals from subsolar-mass black-hole binaries, including systems that may be of primordial origin. The estimated horizon distance extends beyond the Galactic Center and, over part of the parameter space, reaches Andromeda. Our computational estimates indicate that the search is feasible with parallel computing resources and that its performance could be improved further through GPU acceleration.
These results establish NUFFT-based resampling as an efficient foundation for semicoherent searches for long-transient gravitational-wave signals. 

\begin{acknowledgments}
This material is based upon work supported by NSF's LIGO Laboratory which is a major facility fully funded by the National Science Foundation. The authors are grateful for computational resources provided by the LIGO Laboratory and supported by National Science Foundation Grants PHY-0757058 and PHY-0823459.
This work is supported by the Australian Research Council Centre of Excellence for Gravitational Wave Discovery (OzGrav), Project Number CE230100016. 
C.P. acknowledges support from the Italian National Institute for Nuclear Physics (INFN, Istituto Nazionale di Fisica Nucleare), through Commissione Scientifica Nazionale II (CSN2), under the VIRGO-Italia and ET-Italia projects.
O.J.P. is supported by the Spanish Ministerio de Ciencia, Innovacion y Universidades Ramón y Cajal, RYC2023-044489-I funded by MCIN/AEI/10.13039/501100011033 and the FSE+ and cofinanced by the Universitat de les Illes Balears (UIB). This work was supported by UIB with funds from the Programa de Foment de la Recerca i la Innovació de la UIB 2024-2026 (supported by the yearly plan of the Tourist Stay Tax ITS2023-086); the Spanish Agencia Estatal de Investigación grants RED2024-153978-E, RED2024-153735-E, PID2025-170644NA-I00, funded by MICIU/AEI/10.13039/501100011033 and the ERDF/EU; and the Comunitat Autònoma de les Illes Balears through the Conselleria d'Educació i Universitats with funds from the ERDF (SINCO2022/18146).
L.S. is also supported by the Australian Research Council Discovery Early Career Researcher Award, Project Number DE240100206. 
\end{acknowledgments}

\appendix
\section{Resampling and convolutions}
\label{sec:resampling_convolution}
Resampling requires evaluating the signal on a uniformly spaced time grid in the resampled time coordinate. This is achieved by interpolating from the non-uniform samples, with the interpolation rule specifying how neighbouring samples contribute to each point on the resampled grid. Equivalently, the interpolation may be represented as a convolution with an interpolation kernel.

The choice of interpolation kernel determines both the accuracy and computational cost of the resampling procedure. In the frequency domain, a kernel with a narrow spectral width is desirable in order to suppress contributions from aliased spectral replicas. A narrow kernel in the time-domain, however, is preferable because it reduces the number of neighbouring samples that must be evaluated for each interpolated point. These requirements are in tension, analogous to the familiar time-frequency tradeoff in the design of window functions. Prolate spheroidal wave functions provide near-optimal localization under such competing constraints. These wave functions, however, are computationally expensive to evaluate, making practical implementation with them infeasible. Realistic choices of convolution kernel must also account for the computational cost of evaluating the kernel and deconvolution terms, and different kernels have different tradeoffs between complexity, accuracy, and other requirements \cite{greengardAcceleratingNonuniformFast2004, pottsFastFourierTransforms2001, barnett2019}.

\section{StackSlide computational cost}
\label{sec:stackslide_cost}
The average number of power read operations per template, corresponding to the number of coherent time segments spanned by a signal, is $\mathcal{O}(10^3)$. For a bank containing $\mathcal{O}(10^{10})$ templates, this requires $\mathcal{O}(10^{13})$ power reads, corresponding to approximately $20\,\mathrm{TB}$ of memory traffic when the powers are stored in single precision. As representative values, a CPU with dual-channel DDR5 memory can sustain approximately $50\,\mathrm{GB\,s^{-1}}$, while a modern consumer GPU with GDDR6X or GDDR7 memory can sustain approximately $500\,\mathrm{GB\,s^{-1}}$. These correspond to evaluation times of $\mathcal{O}(10^3)\,\mathrm{s}$ on a CPU and $\mathcal{O}(10^2)\,\mathrm{s}$ on a GPU. 

\def\bibsection{\section*{References}}
\bibliographystyle{apsrev4-2-author-truncate.bst}
\bibliography{zotero.bib}

@misc{Alestas:2024ubs,
  title = {Applying the {{Viterbi Algorithm}} to {{Planetary-Mass Black Hole Searches}}},
  year = 2024,
  month = feb,
  number = {arXiv:2401.02314},
  eprint = {2401.02314},
  primaryclass = {astro-ph, physics:gr-qc},
  publisher = {arXiv},
  author = {Alestas, George and Morras, Gonzalo and Yamamoto, Takahiro S. and others}
}

@misc{Andres-Carcasona:2024jvz,
  title = {New Approach to Search for Long Transient Gravitational Waves from Inspiraling Compact Binary Systems},
  year = 2024,
  month = nov,
  number = {arXiv:2411.04498},
  eprint = {2411.04498},
  primaryclass = {gr-qc},
  publisher = {arXiv},
  doi = {10.48550/arXiv.2411.04498},
  author = {Andrés-Carcasona, M. and Piccinni, O. J. and Martínez, M. and others}
}

@article{Arun:2004hn,
  title = {Parameter Estimation of Inspiralling Compact Binaries Using 3.5 Post-{{Newtonian}} Gravitational Wave Phasing: {{The}} Nonspinning Case},
  shorttitle = {Parameter Estimation of Inspiralling Compact Binaries Using 3.5 Post-{{Newtonian}} Gravitational Wave Phasing},
  year = 2005,
  month = apr,
  journal = {Physical Review D},
  volume = {71},
  number = {8},
  pages = {084008},
  publisher = {American Physical Society},
  doi = {10.1103/PhysRevD.71.084008},
  author = {Arun, K. G. and Iyer, Bala R and Sathyaprakash, B. S. and others}
}

@article{Astone:2010ct,
  title = {Data Analysis of Gravitational-Wave Signals from Spinning Neutron Stars. {{V}}. {{A}} Narrow-Band All-Sky Search},
  year = 2010,
  month = jul,
  journal = {Physical Review D},
  volume = {82},
  number = {2},
  pages = {022005},
  issn = {1550-7998, 1550-2368},
  doi = {10.1103/PhysRevD.82.022005},
  author = {Astone, Pia and Borkowski, Kazimierz M. and Jaranowski, Piotr and others}
}

@article{Astone:2014dea,
  title = {Method for Narrow-Band Search of Continuous Gravitational Wave Signals},
  year = 2014,
  month = mar,
  journal = {Physical Review D},
  volume = {89},
  number = {6},
  pages = {062008},
  publisher = {American Physical Society},
  doi = {10.1103/PhysRevD.89.062008},
  author = {Astone, P. and Colla, A. and D’Antonio, S. and others}
}

@article{Astone:2014esa,
  title = {Method for All-Sky Searches of Continuous Gravitational Wave Signals Using the Frequency-{{Hough}} Transform},
  year = 2014,
  month = aug,
  journal = {Physical Review D},
  volume = {90},
  number = {4},
  pages = {042002},
  issn = {1550-7998, 1550-2368},
  doi = {10.1103/PhysRevD.90.042002},
  author = {Astone, Pia and Colla, Alberto and D’Antonio, Sabrina and others}
}

@article{Banagiri:2019obu,
  title = {Search Strategies for Long Gravitational-Wave Transients: {{Hidden Markov}} Model Tracking and Seedless Clustering},
  shorttitle = {Search Strategies for Long Gravitational-Wave Transients},
  year = 2019,
  month = jul,
  journal = {Physical Review D},
  volume = {100},
  number = {2},
  pages = {024034},
  publisher = {American Physical Society},
  doi = {10.1103/PhysRevD.100.024034},
  author = {Banagiri, Sharan and Sun, Ling and Coughlin, Michael W. and others}
}

@article{barnett2019,
  title = {A Parallel Nonuniform Fast Fourier Transform Library Based on an ``Exponential of Semicircle'' Kernel},
  author = {Barnett, Alexander H. and Magland, Jeremy and {af Klinteberg}, Ludvig},
  year = 2019,
  month = jan,
  journal = {SIAM Journal on Scientific Computing},
  volume = {41},
  number = {5},
  eprint = {1808.06736},
  primaryclass = {math.NA},
  pages = {C479-C504},
  doi = {10.1137/18M120885X}
}

@article{Blanchet:2001ax,
  title = {Gravitational-Wave Inspiral of Compact Binary Systems to 7/2 Post-{{Newtonian}} Order},
  year = 2002,
  month = feb,
  journal = {Physical Review D},
  volume = {65},
  number = {6},
  pages = {061501},
  publisher = {American Physical Society},
  doi = {10.1103/PhysRevD.65.061501},
  author = {Blanchet, Luc and Faye, Guillaume and Iyer, Bala R. and others}
}

@article{Blanchet:2023bwj,
  title = {Gravitational-{{Wave Phasing}} of {{Quasicircular Compact Binary Systems}} to the {{Fourth-and-a-Half Post-Newtonian Order}}},
  author = {Blanchet, Luc},
  year = 2023,
  journal = {Physical Review Letters},
  volume = {131},
  number = {12},
  doi = {10.1103/PhysRevLett.131.121402}
}

@article{Boyle:2009dg,
  title = {Comparison of High-Accuracy Numerical Simulations of Black-Hole Binaries with Stationary-Phase Post-{{Newtonian}} Template Waveforms for Initial and Advanced {{LIGO}}},
  author = {Boyle, Michael and Brown, Duncan A and Pekowsky, Larne},
  year = 2009,
  month = may,
  journal = {Classical and Quantum Gravity},
  volume = {26},
  number = {11},
  pages = {114006},
  issn = {0264-9381},
  doi = {10.1088/0264-9381/26/11/114006}
}

@article{Braccini:2011zz,
  title = {Resampling Technique to Correct for the {{Doppler}} Effect in a Search for Gravitational Waves},
  year = 2011,
  month = feb,
  journal = {Physical Review D},
  volume = {83},
  number = {4},
  pages = {044033},
  publisher = {American Physical Society},
  doi = {10.1103/PhysRevD.83.044033},
  author = {Braccini, S. and Cella, G. and Ferrante, I. and others}
}

@article{Brady:1998nj,
  title = {Searching for Periodic Sources with {{LIGO}}. {{II}}. {{Hierarchical}} Searches},
  author = {Brady, Patrick R. and Creighton, Teviet},
  year = 2000,
  month = feb,
  journal = {Physical Review D},
  volume = {61},
  number = {8},
  pages = {082001},
  issn = {0556-2821, 1089-4918},
  doi = {10.1103/PhysRevD.61.082001},
  copyright = {http://link.aps.org/licenses/aps-default-license}
}

@article{Brown:2012nn,
  title = {Template Banks to Search for Low-Mass Binary Black Holes in Advanced Gravitational-Wave Detectors},
  author = {Brown, Duncan A. and Kumar, Prayush and Nitz, Alexander H.},
  year = 2013,
  month = apr,
  journal = {Physical Review D},
  volume = {87},
  number = {8},
  pages = {082004},
  publisher = {American Physical Society},
  doi = {10.1103/PhysRevD.87.082004}
}

@article{Buonanno:2009zt,
  title = {Comparison of Post-{{Newtonian}} Templates for Compact Binary Inspiral Signals in Gravitational-Wave Detectors},
  year = 2009,
  month = oct,
  journal = {Physical Review D},
  volume = {80},
  number = {8},
  pages = {084043},
  issn = {1550-7998, 1550-2368},
  doi = {10.1103/PhysRevD.80.084043},
  copyright = {http://link.aps.org/licenses/aps-default-license},
  author = {Buonanno, Alessandra and Iyer, Bala R. and Ochsner, Evan and others}
}

@article{Capote:2024rmo,
  title = {Advanced {{LIGO}} Detector Performance in the Fourth Observing Run},
  year = 2025,
  month = mar,
  journal = {Physical Review D},
  volume = {111},
  number = {6},
  pages = {062002},
  issn = {2470-0010, 2470-0029},
  doi = {10.1103/PhysRevD.111.062002},
  author = {Capote, E. and Jia, W. and Aritomi, N. and others}
}

@article{Carr:1975qj,
  title = {The Primordial Black Hole Mass Spectrum.},
  author = {Carr, B. J.},
  year = 1975,
  journal = {The Astrophysical Journal},
  month = oct,
  volume = {201},
  pages = {1--19},
  doi = {10.1086/153853}
}

@article{Carr:2020xqk,
  title = {Primordial {{Black Holes}} as {{Dark Matter}}: {{Recent Developments}}},
  shorttitle = {Primordial {{Black Holes}} as {{Dark Matter}}},
  author = {Carr, Bernard and Kuhnel, Florian},
  year = 2020,
  month = oct,
  journal = {Annual Review of Nuclear and Particle Science},
  volume = {70},
  number = {1},
  eprint = {2006.02838},
  primaryclass = {astro-ph, physics:gr-qc, physics:hep-th},
  pages = {355--394},
  issn = {0163-8998, 1545-4134},
  doi = {10.1146/annurev-nucl-050520-125911}
}

@article{Carr:2023tpt,
  title = {Observational Evidence for Primordial Black Holes: {{A}} Positivist Perspective},
  shorttitle = {Observational Evidence for Primordial Black Holes},
  year = 2024,
  month = feb,
  journal = {Physics Reports},
  volume = {1054},
  pages = {1--68},
  issn = {03701573},
  doi = {10.1016/j.physrep.2023.11.005},
  author = {Carr, B.J. and Clesse, S. and García-Bellido, J. and others}
}

@article{duttFastFourierTransforms1993,
  title = {Fast {{Fourier Transforms For Nonequispaced Data}}},
  author = {Dutt, A and Rokhlin, V.},
  year = 1993,
  month = nov,
  journal = {SIAM Journal on Scientific Computing},
  volume = {14},
  number = {6},
  pages = {1368--1393},
  doi = {10.1137/0914081}
}

@article{Faye:2012we,
  title = {The Third and a Half-Post-{{Newtonian}} Gravitational Wave Quadrupole Mode for Quasi-Circular Inspiralling Compact Binaries},
  year = 2012,
  month = aug,
  journal = {Classical and Quantum Gravity},
  volume = {29},
  number = {17},
  pages = {175004},
  publisher = {IOP Publishing},
  issn = {0264-9381},
  doi = {10.1088/0264-9381/29/17/175004},
  author = {Faye, Guillaume and Marsat, Sylvain and Blanchet, Luc and others}
}

@misc{finufftFlatironinstituteFinufftNonuniform,
  title = {Flatironinstitute/Finufft: {{Non-uniform}} Fast {{Fourier}} Transform Library of Types 1,2,3 in Dimensions 1,2,3},
  author = {FINUFFT},
  howpublished = {https://github.com/flatironinstitute/finufft/tree/master}
}

@article{Grace:2023kqq,
  title = {Piecewise Frequency Model for Searches for Long-Transient Gravitational Waves from Young Neutron Stars},
  year = 2023,
  month = dec,
  journal = {Physical Review D},
  volume = {108},
  number = {12},
  pages = {123045},
  publisher = {American Physical Society},
  doi = {10.1103/PhysRevD.108.123045},
  author = {Grace, Benjamin and Wette, Karl and Scott, Susan M. and others}
}

@article{Green:2020jor,
  title = {Primordial Black Holes as a Dark Matter Candidate},
  author = {Green, Anne M and Kavanagh, Bradley J},
  year = 2021,
  month = apr,
  journal = {Journal of Physics G: Nuclear and Particle Physics},
  volume = {48},
  number = {4},
  pages = {043001},
  issn = {0954-3899, 1361-6471},
  doi = {10.1088/1361-6471/abc534}
}

@article{greengardAcceleratingNonuniformFast2004,
  title = {Accelerating the {{Nonuniform Fast Fourier Transform}}},
  author = {Greengard, Leslie and Lee, June-Yub},
  year = 2004,
  month = jan,
  journal = {SIAM Review},
  volume = {46},
  number = {3},
  pages = {443--454},
  issn = {0036-1445, 1095-7200},
  doi = {10.1137/S003614450343200X}
}

@misc{Hanna:2022zpk,
  title = {A Binary Tree Approach to Template Placement for Searches for Gravitational Waves from Compact Binary Mergers},
  year = 2022,
  month = sep,
  number = {arXiv:2209.11298},
  eprint = {2209.11298},
  primaryclass = {gr-qc},
  publisher = {arXiv},
  doi = {10.48550/arXiv.2209.11298},
  author = {Hanna, Chad and Kennington, James and Sakon, Shio and others}
}

@article{Hanna:2024tom,
  title = {Template Bank for Subsolar Mass Compact Binary Mergers in the Fourth Observing Run of {{Advanced LIGO}}, {{Advanced Virgo}}, and {{KAGRA}}},
  year = 2025,
  month = aug,
  journal = {Physical Review D},
  volume = {112},
  number = {4},
  pages = {044013},
  publisher = {American Physical Society},
  doi = {10.1103/c97v-bmj8},
  author = {Hanna, Chad and Kennington, James and Niu, Wanting and others}
}

@article{Haskell:2023exob,
  title = {Glitching Pulsars as Gravitational Wave Sources},
  author = {Haskell, B. and Jones, D. I.},
  year = 2024,
  month = may,
  journal = {Astroparticle Physics},
  volume = {157},
  pages = {102921},
  issn = {0927-6505},
  doi = {10.1016/j.astropartphys.2023.102921}
}

@article{Hawking:1971ei,
  title = {Gravitationally {{Collapsed Objects}} of {{Very Low Mass}}},
  author = {Hawking, Stephen},
  year = 1971,
  month = apr,
  journal = {Monthly Notices of the Royal Astronomical Society},
  volume = {152},
  number = {1},
  pages = {75--78},
  issn = {0035-8711},
  doi = {10.1093/mnras/152.1.75}
}

@article{Jaranowski:1998qm,
  title = {Data Analysis of Gravitational-Wave Signals from Spinning Neutron Stars: {{The}} Signal and Its Detection},
  shorttitle = {Data Analysis of Gravitational-Wave Signals from Spinning Neutron Stars},
  author = {Jaranowski, Piotr and Kr{\'o}lak, Andrzej and Schutz, Bernard F.},
  year = 1998,
  month = aug,
  journal = {Physical Review D},
  volume = {58},
  number = {6},
  pages = {063001},
  issn = {0556-2821, 1089-4918},
  doi = {10.1103/PhysRevD.58.063001}
}

@article{Jaranowski:2005hz,
  title = {Gravitational-{{Wave Data Analysis}}. {{Formalism}} and {{Sample Applications}}: {{The Gaussian Case}}},
  shorttitle = {Gravitational-{{Wave Data Analysis}}. {{Formalism}} and {{Sample Applications}}},
  author = {Jaranowski, Piotr and Kr{\'o}lak, Andrzej},
  year = 2012,
  month = mar,
  journal = {Living Reviews in Relativity},
  volume = {15},
  number = {1},
  pages = {4},
  issn = {1433-8351},
  doi = {10.12942/lrr-2012-4}
}

@article{Jones:2023fzz,
  title = {Methods and Prospects for Gravitational-Wave Searches Targeting Ultralight Vector-Boson Clouds around Known Black Holes},
  year = 2023,
  month = sep,
  journal = {Physical Review D},
  volume = {108},
  number = {6},
  pages = {064001},
  publisher = {American Physical Society},
  doi = {10.1103/PhysRevD.108.064001},
  author = {Jones, Dana and Sun, Ling and Siemonsen, Nils and others}
}

@article{Kacanja:2024pjh,
  title = {Efficient {{Stochastic Template Bank Using Inner Product Inequalities}}},
  year = 2024,
  month = nov,
  journal = {The Astrophysical Journal},
  volume = {975},
  number = {2},
  pages = {212},
  publisher = {The American Astronomical Society},
  issn = {0004-637X},
  doi = {10.3847/1538-4357/ad7d87},
  author = {Kacanja, Keisi and Nitz, Alexander H. and Wu, Shichao and others}
}

@misc{Kacanja:2026byy,
  title = {Search for {{Sub-Solar Mass Binaries}} in the {{First Part}} of {{LIGO}}'s {{Fourth Observing Run}}},
  year = 2026,
  month = mar,
  number = {arXiv:2602.12115},
  eprint = {2602.12115},
  primaryclass = {astro-ph.HE},
  publisher = {arXiv},
  doi = {10.48550/arXiv.2602.12115},
  author = {Kacanja, Keisi and Soni, Kanchan and Akyüz, Aleyna and others}
}

@article{KAGRA:2020tym,
  title = {Overview of {{KAGRA}}: {{Detector}} Design and Construction History},
  year = 2020,
  month = aug,
  journal = {Progress of Theoretical and Experimental Physics},
  volume = {2021},
  number = {5},
  eprint = {https://academic.oup.com/ptep/article-pdf/2021/5/05A101/37974994/ptaa125.pdf},
  pages = {05A101},
  issn = {2050-3911},
  doi = {10.1093/ptep/ptaa125},
  author = {Akutsu, T and Ando, M and Arai, K and others}
}

@article{Keitel:2019zhb,
  title = {First Search for Long-Duration Transient Gravitational Waves after Glitches in the {{Vela}} and {{Crab}} Pulsars},
  year = 2019,
  month = sep,
  journal = {Physical Review D},
  volume = {100},
  number = {6},
  pages = {064058},
  publisher = {American Physical Society},
  doi = {10.1103/PhysRevD.100.064058},
  author = {Keitel, David and Woan, Graham and Pitkin, Matthew and others}
}

@article{Krishnan:2004sv,
  title = {Hough Transform Search for Continuous Gravitational Waves},
  year = 2004,
  month = oct,
  journal = {Physical Review D},
  volume = {70},
  number = {8},
  pages = {082001},
  issn = {1550-7998, 1550-2368},
  doi = {10.1103/PhysRevD.70.082001},
  author = {Krishnan, Badri and Sintes, Alicia M. and Papa, Maria Alessandra and others}
}

@misc{lalsuite,
  title = {{{LVK Algorithm Library}} - {{LALSuite}}},
  author = {{LIGO Scientific Collaboration} and {Virgo Collaboration} and {KAGRA Collaboration}},
  year = 2018,
  doi = {10.7935/GT1W-FZ16},
  howpublished = {Free software (GPL)}
}

@article{LIGOScientific:2014pky,
  title = {Advanced {{LIGO}}},
  year = 2015,
  month = mar,
  journal = {Classical and Quantum Gravity},
  volume = {32},
  number = {7},
  pages = {074001},
  publisher = {IOP Publishing},
  doi = {10.1088/0264-9381/32/7/074001},
  author = {Aasi, J and Abbott, B P and Abbott, R and others}
}

@article{LIGOScientific:2016aoc,
  title = {Observation of {{Gravitational Waves}} from a {{Binary Black Hole Merger}}},
  year = 2016,
  month = feb,
  journal = {Physical Review Letters},
  volume = {116},
  number = {6},
  pages = {061102},
  publisher = {American Physical Society},
  doi = {10.1103/PhysRevLett.116.061102},
  author = {Abbott, B. P. and Abbott, R. and Abbott, T. D. and others}
}

@article{LIGOScientific:2017fdd,
  title = {Search for {{Post-merger Gravitational Waves}} from the {{Remnant}} of the {{Binary Neutron Star Merger GW170817}}},
  year = 2017,
  month = dec,
  journal = {The Astrophysical Journal Letters},
  volume = {851},
  number = {1},
  pages = {L16},
  publisher = {The American Astronomical Society},
  issn = {2041-8205},
  doi = {10.3847/2041-8213/aa9a35},
  author = {Abbott, B. P. and Abbott, R. and Abbott, T. D. and others}
}

@article{LIGOScientific:2017ytx,
  title = {First Narrow-Band Search for Continuous Gravitational Waves from Known Pulsars in Advanced Detector Data},
  year = 2017,
  month = dec,
  journal = {Physical Review D},
  volume = {96},
  number = {12},
  pages = {122006},
  publisher = {American Physical Society},
  doi = {10.1103/PhysRevD.96.122006},
  author = {Abbott, B. P. and Abbott, R. and Abbott, T. D. and others}
}

@misc{LIGOScientific:2025csr,
  title = {Directed Searches for Gravitational Waves from Ultralight Vector Boson Clouds around Merger Remnant and Galactic Black Holes during the First Part of the Fourth {{LIGO-Virgo-KAGRA}} Observing Run},
  year = 2026,
  eprint = {2509.07352},
  primaryclass = {gr-qc},
  author = {Abac, A. G. and Abouelfettouh, I. and Acernese, F. and others}
}

@article{LIGOScientific:2025slb,
  title = {{{GWTC-4}}.0: {{Updating}} the {{Gravitational-Wave Transient Catalog}} with {{Observations}} from the {{First Part}} of the {{Fourth LIGO-Virgo-KAGRA Observing Run}}},
  shorttitle = {{{GWTC-4}}.0},
  year = 2026,
  month = jun,
  journal = {The Astrophysical Journal Letters},
  volume = {1004},
  number = {2},
  eprint = {2508.18082},
  primaryclass = {gr-qc},
  pages = {L22},
  issn = {2041-8205, 2041-8213},
  doi = {10.3847/2041-8213/ae2c74},
  author = {Abac, A. G. and Abouelfettouh, I. and Acernese, F. and others}
}

@misc{LIGOScientific:2025snk,
  title = {Open {{Data}} from {{LIGO}}, {{Virgo}}, and {{KAGRA}} through the {{First Part}} of the {{Fourth Observing Run}}},
  year = 2025,
  month = nov,
  number = {arXiv:2508.18079},
  eprint = {2508.18079},
  primaryclass = {gr-qc},
  publisher = {arXiv},
  doi = {10.48550/arXiv.2508.18079},
  author = {Abac, A. G. and Abouelfettouh, I. and Acernese, F. and others}
}

@article{LIGOScientific:2026qsb,
  title = {Narrowband {{Searches}} for {{Continuous Gravitational Waves}} from Known {{Pulsars}} in the {{First Two Parts}} of the {{Fourth LIGO}}--{{Virgo}}--{{KAGRA Observing Run}}},
  year = 2026,
  month = jul,
  journal = {The Astrophysical Journal},
  volume = {1005},
  number = {2},
  pages = {221},
  publisher = {The American Astronomical Society},
  issn = {0004-637X},
  doi = {10.3847/1538-4357/ae77fc},
  author = {Abac, A. G. and Abouelfettouh, I. and Acernese, F. and others}
}

@misc{LIGOScientific:2026wfs,
  title = {{{GWTC-5}}.0: {{Observations}} from the {{Second Part}} of the {{Fourth LIGO-Virgo-KAGRA Observing Run}} and {{Updates}} to the {{Gravitational-Wave Transient Catalog}}},
  shorttitle = {{{GWTC-5}}.0},
  year = 2026,
  month = jun,
  number = {arXiv:2605.27225},
  eprint = {2605.27225},
  primaryclass = {gr-qc},
  publisher = {arXiv},
  doi = {10.48550/arXiv.2605.27225},
  author = {Abac, A. G. and Abe, A. and Abouelfettouh, I. and others}
}

@misc{LIGOScientific:2026wxz,
  title = {Searches for {{Binary Mergers}} with {{Sub-solar Mass Components}} in {{Data}} from the {{First Part}} of {{LIGO--Virgo--KAGRA}}'s {{Fourth Observing Run}}},
  year = 2026,
  month = may,
  number = {arXiv:2605.05444},
  eprint = {2605.05444},
  primaryclass = {astro-ph.HE},
  publisher = {arXiv},
  doi = {10.48550/arXiv.2605.05444},
  author = {Abac, A. G. and Abouelfettouh, I. and Acernese, F. and others}
}

@misc{LIGOT2000012v2NoiseCurves,
  title = {{{LIGO-T2000012-v2}}: {{Noise}} Curves Used for {{Simulations}} in the Update of the {{Observing Scenarios Paper}}},
  howpublished = {https://dcc.ligo.org/LIGO-T2000012/public}
}

@misc{linJyhmiinlinPynufft2023,
  title = {Jyhmiinlin/Pynufft},
  author = {Lin, Jyh-Miin},
  year = 2023,
  month = nov,
  howpublished = {https://github.com/jyhmiinlin/pynufft}
}

@article{LISACosmologyWorkingGroup:2023njw,
  title = {Primordial Black Holes and Their Gravitational-Wave Signatures},
  year = 2025,
  month = jan,
  journal = {Living Reviews in Relativity},
  volume = {28},
  number = {1},
  pages = {1},
  issn = {1433-8351},
  doi = {10.1007/s41114-024-00053-w},
  author = {Bagui, Eleni and Clesse, Sébastien and De Luca, Valerio and others}
}

@incollection{livasBroadbandSearchTechniques1989,
  title = {Broadband Search Techniques for Periodic Sources of Gravitational Radiation},
  booktitle = {Gravitational Wave Data Analysis},
  author = {Livas, Jeffrey},
  editor = {Schutz, B. F.},
  year = 1989,
  pages = {217--238},
  publisher = {Springer Netherlands},
  address = {Dordrecht},
  doi = {10.1007/978-94-009-1185-7_16},
  isbn = {978-94-009-1185-7}
}

@article{LVK:2022ydq,
  title = {Search for Subsolar-Mass Black Hole Binaries in the Second Part of {{Advanced LIGO}}'s and {{Advanced Virgo}}'s Third Observing Run},
  author = {{The LVK Collaboration}},
  year = 2023,
  month = jul,
  journal = {Monthly Notices of the Royal Astronomical Society},
  volume = {524},
  number = {4},
  pages = {5984--5992},
  issn = {0035-8711, 1365-2966},
  doi = {10.1093/mnras/stad588}
}

@article{Macleod:2021goi,
  title = {{{GWpy}}: {{A Python}} Package for Gravitational-Wave Astrophysics},
  shorttitle = {{{GWpy}}},
  year = 2021,
  month = jan,
  journal = {SoftwareX},
  volume = {13},
  pages = {100657},
  issn = {2352-7110},
  doi = {10.1016/j.softx.2021.100657},
  author = {Macleod, Duncan M. and Areeda, Joseph S. and Coughlin, Scott B. and others}
}

@article{Macquet:2021ttqa,
  title = {Long-Duration Transient Gravitational-Wave Search Pipeline},
  year = 2021,
  month = nov,
  journal = {Physical Review D},
  volume = {104},
  number = {10},
  pages = {102005},
  issn = {2470-0010, 2470-0029},
  doi = {10.1103/PhysRevD.104.102005},
  author = {Macquet, A. and Bizouard, M. A. and Christensen, N. and others}
}

@article{Mastrogiovanni:2017xjr,
  title = {An Improved Algorithm for Narrow-Band Searches of Continuous Gravitational Waves},
  year = 2017,
  month = jul,
  journal = {Classical and Quantum Gravity},
  volume = {34},
  number = {13},
  eprint = {1703.03493},
  primaryclass = {gr-qc},
  pages = {135007},
  issn = {0264-9381, 1361-6382},
  doi = {10.1088/1361-6382/aa744f},
  author = {Mastrogiovanni, S. and Astone, P. and D'Antonio, S. and others}
}

@article{Meadors:2017pefa,
  title = {Resampling to Accelerate Cross-Correlation Searches for Continuous Gravitational Waves from Binary Systems},
  year = 2018,
  month = feb,
  journal = {Physical Review D},
  volume = {97},
  number = {4},
  pages = {044017},
  publisher = {American Physical Society},
  doi = {10.1103/PhysRevD.97.044017},
  author = {Meadors, Grant David and Krishnan, Badri and Papa, Maria Alessandra and others}
}

@article{Mendell_stackslide_hough,
  title = {{{StackSlide}} and {{Hough Search SNR}} and {{Statistics}}},
  author = {Mendell, Gregory and Landry, Michael},
  year = 2005,
  journal = {Tech. Rep. T050003-x0, LIGO}
}

@article{Miller:2018rbg,
  title = {A Method to Search for Long Duration Gravitational Wave Transients from Isolated Neutron Stars Using the Generalized {{FrequencyHough}}},
  year = 2018,
  month = nov,
  journal = {Physical Review D},
  volume = {98},
  number = {10},
  eprint = {1810.09784},
  primaryclass = {astro-ph, physics:physics},
  pages = {102004},
  issn = {2470-0010, 2470-0029},
  doi = {10.1103/PhysRevD.98.102004},
  author = {Miller, Andrew and Astone, Pia and D'Antonio, Sabrina and others}
}

@article{Miller:2024rca,
  title = {Gravitational Waves from Sub-Solar Mass Primordial Black Holes},
  author = {Miller, Andrew L.},
  year = 2024,
  month = apr,
  journal = {arXiv e-prints},
  number = {arXiv:2404.11601},
  eprint = {2404.11601},
  primaryclass = {gr-qc},
  pages = {arXiv:2404.11601},
  doi = {10.48550/arXiv.2404.11601}
}

@phdthesis{ouGNUFFTWAutoTuningHighPerformance2017,
  title = {{{gNUFFTW}}: {{Auto-Tuning}} for {{High-Performance GPU-Accelerated Non-Uniform Fast Fourier Transforms}}},
  author = {Ou, Teresa},
  year = 2017,
  school = {EECS Department, University of California, Berkeley}
}

@article{Owen:1998dk,
  title = {Matched Filtering of Gravitational Waves from Inspiraling Compact Binaries: {{Computational}} Cost and Template Placement},
  shorttitle = {Matched Filtering of Gravitational Waves from Inspiraling Compact Binaries},
  author = {Owen, Benjamin J. and Sathyaprakash, B. S.},
  year = 1999,
  month = jun,
  journal = {Physical Review D},
  volume = {60},
  number = {2},
  pages = {022002},
  publisher = {American Physical Society},
  doi = {10.1103/PhysRevD.60.022002}
}

@article{Pan:2007nw,
  title = {Data-Analysis Driven Comparison of Analytic and Numerical Coalescing Binary Waveforms: {{Nonspinning}} Case},
  shorttitle = {Data-Analysis Driven Comparison of Analytic and Numerical Coalescing Binary Waveforms},
  year = 2008,
  month = jan,
  journal = {Physical Review D},
  volume = {77},
  number = {2},
  pages = {024014},
  publisher = {American Physical Society},
  doi = {10.1103/PhysRevD.77.024014},
  author = {Pan, Yi and Buonanno, Alessandra and Baker, John G. and others}
}

@article{Piccinni:2018akm,
  title = {A New Data Analysis Framework for the Search of Continuous Gravitational Wave Signals},
  year = 2019,
  month = jan,
  journal = {Classical and Quantum Gravity},
  volume = {36},
  number = {1},
  eprint = {1811.04730},
  primaryclass = {gr-qc},
  pages = {015008},
  issn = {0264-9381, 1361-6382},
  doi = {10.1088/1361-6382/aaefb5},
  author = {Piccinni, O. J. and Frasca, S. and Astone, P. and others}
}

@incollection{pottsFastFourierTransforms2001,
  title = {Fast {{Fourier Transforms}} for {{Nonequispaced Data}}: {{A Tutorial}}},
  shorttitle = {Fast {{Fourier Transforms}} for {{Nonequispaced Data}}},
  booktitle = {Modern {{Sampling Theory}}},
  author = {Potts, Daniel and Steidl, Gabriele and Tasche, Manfred},
  editor = {Benedetto, John J. and Ferreira, Paulo J. S. G.},
  year = 2001,
  pages = {247--270},
  publisher = {Birkh\"auser Boston},
  address = {Boston, MA},
  doi = {10.1007/978-1-4612-0143-4_12},
  isbn = {978-1-4612-6632-7 978-1-4612-0143-4}
}

@article{Prix:2007ks,
  title = {Template-Based Searches for Gravitational Waves: Efficient Lattice Covering of Flat Parameter Spaces},
  shorttitle = {Template-Based Searches for Gravitational Waves},
  author = {Prix, Reinhard},
  year = 2007,
  month = oct,
  journal = {Classical and Quantum Gravity},
  volume = {24},
  number = {19},
  eprint = {0707.0428},
  primaryclass = {gr-qc},
  pages = {S481-S490},
  issn = {0264-9381, 1361-6382},
  doi = {10.1088/0264-9381/24/19/S11}
}

@article{Prix:2011qv,
  title = {Search Method for Long-Duration Gravitational-Wave Transients from Neutron Stars},
  author = {Prix, R. and Giampanis, S. and Messenger, C.},
  year = 2011,
  month = jul,
  journal = {Physical Review D},
  volume = {84},
  number = {2},
  pages = {023007},
  issn = {1550-7998, 1550-2368},
  doi = {10.1103/PhysRevD.84.023007},
  copyright = {http://link.aps.org/licenses/aps-default-license}
}

@misc{Rodriguez:2026oyia,
  title = {Search for {{Planetary-mass Black Holes}} with an {{Improved Viterbi Algorithm}}},
  year = 2026,
  month = jul,
  number = {arXiv:2607.18352},
  eprint = {2607.18352},
  primaryclass = {astro-ph.IM},
  publisher = {arXiv},
  doi = {10.48550/arXiv.2607.18352},
  author = {Rodriguez, Raul and Alestas, George and Kuroyanagi, Sachiko and others}
}

@article{Roy:2017oul,
  title = {Effectual Template Banks for Upcoming Compact Binary Searches in {{Advanced-LIGO}} and {{Virgo}} Data},
  author = {Roy, Soumen and Sengupta, Anand S. and Ajith, Parameswaran},
  year = 2019,
  month = jan,
  journal = {Physical Review D},
  volume = {99},
  number = {2},
  pages = {024048},
  publisher = {American Physical Society},
  doi = {10.1103/PhysRevD.99.024048}
}

@article{Roy:2017qgg,
  title = {Hybrid Geometric-Random Template-Placement Algorithm for Gravitational Wave Searches from Compact Binary Coalescences},
  author = {Roy, Soumen and Sengupta, Anand S. and Thakor, Nilay},
  year = 2017,
  month = may,
  journal = {Physical Review D},
  volume = {95},
  number = {10},
  pages = {104045},
  publisher = {American Physical Society},
  doi = {10.1103/PhysRevD.95.104045}
}

@misc{shihCuFINUFFTLoadbalancedGPU2021,
  title = {{{cuFINUFFT}}: A Load-Balanced {{GPU}} Library for General-Purpose Nonuniform {{FFTs}}},
  shorttitle = {{{cuFINUFFT}}},
  year = 2021,
  month = mar,
  number = {arXiv:2102.08463},
  eprint = {2102.08463},
  primaryclass = {cs, eess, math},
  publisher = {arXiv},
  author = {Shih, Yu-hsuan and Wright, Garrett and Andén, Joakim and others}
}

@article{Siemonsen:2019ebd,
  title = {Gravitational Wave Signatures of Ultralight Vector Bosons from Black Hole Superradiance},
  author = {Siemonsen, Nils and East, William E.},
  year = 2020,
  month = jan,
  journal = {Physical Review D},
  volume = {101},
  number = {2},
  pages = {024019},
  issn = {2470-0010, 2470-0029},
  doi = {10.1103/PhysRevD.101.024019}
}

@article{Singhal:2019dfn,
  title = {A Resampling Algorithm to Detect Continuous Gravitational-Wave Signals from Neutron Stars in Binary Systems},
  year = 2019,
  month = oct,
  journal = {Classical and Quantum Gravity},
  volume = {36},
  number = {20},
  pages = {205015},
  issn = {0264-9381, 1361-6382},
  doi = {10.1088/1361-6382/ab4367},
  author = {Singhal, A and Leaci, P and Astone, P and others}
}

@article{Smith:1987hh,
  title = {Algorithm to Search for Gravitational Radiation from Coalescing Binaries},
  author = {Smith, Sheryl},
  year = 1987,
  month = nov,
  journal = {Physical Review D},
  volume = {36},
  number = {10},
  pages = {2901--2904},
  issn = {0556-2821},
  doi = {10.1103/PhysRevD.36.2901}
}

@article{Sun:2017zge,
  title = {Hidden {{Markov}} Model Tracking of Continuous Gravitational Waves from Young Supernova Remnants},
  year = 2018,
  month = feb,
  journal = {Physical Review D},
  volume = {97},
  number = {4},
  pages = {043013},
  issn = {2470-0010, 2470-0029},
  doi = {10.1103/PhysRevD.97.043013},
  author = {Sun, L. and Melatos, A. and Suvorova, S. and others}
}

@article{Sun:2018owi,
  title = {Application of Hidden {{Markov}} Model Tracking to the Search for Long-Duration Transient Gravitational Waves from the Remnant of the Binary Neutron Star Merger {{GW170817}}},
  author = {Sun, Ling and Melatos, Andrew},
  year = 2019,
  month = jun,
  journal = {Physical Review D},
  volume = {99},
  number = {12},
  pages = {123003},
  publisher = {American Physical Society},
  doi = {10.1103/PhysRevD.99.123003}
}

@article{Suvorova:2016rdc,
  title = {Hidden {{Markov}} Model Tracking of Continuous Gravitational Waves from a Neutron Star with Wandering Spin},
  year = 2016,
  month = jun,
  journal = {Physical Review D},
  volume = {93},
  number = {12},
  pages = {123009},
  issn = {2470-0010, 2470-0029},
  doi = {10.1103/PhysRevD.93.123009},
  copyright = {http://link.aps.org/licenses/aps-default-license},
  author = {Suvorova, S. and Sun, L. and Melatos, A. and others}
}

@article{swiglal,
  title = {{{SWIGLAL}}: {{Python}} and {{Octave}} Interfaces to the {{LALSuite}} Gravitational-Wave Data Analysis Libraries\textbraceright},
  author = {Wette, Karl},
  year = 2020,
  journal = {SoftwareX},
  volume = {12},
  pages = {100634},
  doi = {10.1016/j.softx.2020.100634}
}

@article{tafti2006,
  title = {On {{Interpolation}} and {{Resampling}} of {{Discrete Data}}},
  author = {Tafti, Pouya Dehghani and Shirani, Shahram and Wu, Xiaolin},
  year = 2006,
  month = dec,
  journal = {IEEE Signal Processing Letters},
  volume = {13},
  number = {12},
  pages = {733--736},
  issn = {1070-9908},
  doi = {10.1109/LSP.2006.879854},
  copyright = {https://ieeexplore.ieee.org/Xplorehelp/downloads/license-information/IEEE.html}
}

@article{Tenorio:2021wmz,
  title = {Search {{Methods}} for {{Continuous Gravitational-Wave Signals}} from {{Unknown Sources}} in the {{Advanced-Detector Era}}},
  author = {Tenorio, Rodrigo and Keitel, David and Sintes, Alicia M.},
  year = 2021,
  month = dec,
  journal = {Universe},
  volume = {7},
  number = {12},
  pages = {474},
  publisher = {Multidisciplinary Digital Publishing Institute},
  issn = {2218-1997},
  doi = {10.3390/universe7120474},
  copyright = {http://creativecommons.org/licenses/by/3.0/}
}

@misc{vaillantGhisvailPyNFFT2023,
  title = {Ghisvail/{{pyNFFT}}},
  author = {Vaillant, Ghislain},
  year = 2023,
  month = nov,
  copyright = {GPL-3.0},
  howpublished = {https://github.com/ghisvail/pyNFFT}
}

@misc{vanderplasJakevdpNfft2023,
  title = {Jakevdp/Nfft},
  author = {Vanderplas, Jake},
  year = 2023,
  month = dec,
  copyright = {MIT},
  howpublished = {https://github.com/jakevdp/nfft}
}

@article{VIRGO:2014yos,
  title = {Advanced {{Virgo}}: A Second-Generation Interferometric Gravitational Wave Detector},
  year = 2014,
  month = dec,
  journal = {Classical and Quantum Gravity},
  volume = {32},
  number = {2},
  pages = {024001},
  publisher = {IOP Publishing},
  doi = {10.1088/0264-9381/32/2/024001},
  author = {Acernese, F and Agathos, M and Agatsuma, K and others}
}

@article{wareFastApproximateFourier1998,
  title = {Fast {{Approximate Fourier Transforms}} for {{Irregularly Spaced Data}}},
  author = {Ware, Antony F.},
  year = 1998,
  month = jan,
  journal = {SIAM Review},
  volume = {40},
  number = {4},
  pages = {838--856},
  issn = {0036-1445, 1095-7200},
  doi = {10.1137/S003614459731533X}
}

@article{Wette:2014tca,
  title = {Lattice Template Placement for Coherent All-Sky Searches for Gravitational-Wave Pulsars},
  author = {Wette, Karl},
  year = 2014,
  month = dec,
  journal = {Physical Review D},
  volume = {90},
  number = {12},
  pages = {122010},
  issn = {1550-7998, 1550-2368},
  doi = {10.1103/PhysRevD.90.122010},
  copyright = {http://link.aps.org/licenses/aps-default-license}
}

@article{Wette:2023dom,
  title = {Searches for Continuous Gravitational Waves from Neutron Stars: {{A}} Twenty-Year Retrospective},
  shorttitle = {Searches for Continuous Gravitational Waves from Neutron Stars},
  author = {Wette, Karl},
  year = 2023,
  month = nov,
  journal = {Astroparticle Physics},
  volume = {153},
  eprint = {2305.07106},
  primaryclass = {astro-ph, physics:gr-qc},
  pages = {102880},
  issn = {09276505},
  doi = {10.1016/j.astropartphys.2023.102880}
}

\end{document}